\documentclass[12pt]{article}
\usepackage{amssymb}
\usepackage{amsmath}

\usepackage{graphicx}

\begin{document}

\begin{center}
\textbf{Supplementary Conjugated Circuits for Biphenylene and related
hydrocarbons}

\textbf{\ Viktorija Gineityte\bigskip }

Institute of Theoretical Physics and Astronomy, Vilnius University, Gostauto
12, LT-01108 Vilnius, Lithuania

Viktorija.Gineityte@tfai.vu.lt\bigskip

\textbf{Abstract}
\end{center}

Individual Kekul\'{e} valence structures of biphenylene and related
hydrocarbons are comparatively studied in respect of their total pi-electron
energies and thereby relative stabilities. These structures are modelled as
sets of weakly-interacting initially-double (C=C) bonds. The relevant total
energies are represented in the form of power series, wherein the averaged
resonance parameter of initially-single (C$-$C) bonds underlies the
expansion. To rationalize the resulting distinctions in total energies,
interrelations are sought between separate members of the series, on the one
hand, and presence of definite substructures in the given \ Kekul\'{e}
valence structure, on the other hand. It is shown that monocycles S$_{1}$
and S$_{2}$ correspondingly containing two and four exocyclic methylene
groups [like 3,4-dimethylene cyclobutene and [4]radialene] participate in
the formation of energy corrections of the relevant Kekul\'{e} valence
structures along with the usual conjugated circuits of the $4n+2$ and $4n$
series (R$_{n}$ and Q$_{n},n=1,2,3...$). Thus, the cycles S$_{1}$ and S$_{2}$%
\ are deductively predicted to play the role of supplementary conjugated
circuits for biphenylene-like hydrocarbons. Moreover, the S$_{2}-$ and S$%
_{1}-$ containing structures are shown to be the most stable ones among all
Kekul\'{e} valence structures of the given hydrocarbon. Meanwhile, the
lowest stability is predicted for structures in which either the neighboring
hexagonal rings are connected by two C=C bonds or two exocyclic C=C bonds
are attached to the same hexagonal ring.

\begin{center}
\textbf{1. Introduction}
\end{center}

Stabilities of pi-electron systems of monocyclic hydrocarbons (annulenes)
are known to be governed by the H\"{u}ckel ($4n+2$) rule (see e.g. [1, 2]),
which predicts an increased stability (and thereby aromaticity) of compounds
having 6, 10, 14,.. carbon atoms and/or pi-electrons vs. those of the 4n
series, viz. C$_{4}$H$_{4},$ C$_{8}$H$_{8}$, etc. For polycyclic molecules,
however, such a simple and universal rule has been not yet formulated in
spite of numerous attempts in this direction. Particular rules based on
total numbers either of Kekul\'{e} valence structures of the given compound
(cf. the Kekul\'{e} structure count [3-6]) or of the aromatic sextets (cf.
the Clar aromatic sextet (CAS) model [7-9]) may be referred to here as the
most well-known examples of early qualitative approaches to stabilities of
polycyclic hydrocarbons. Later and more sophisticated treatments of the same
problem, in turn, may be exemplified by the theory of cyclic conjugation
[10] and by that of conjugated circuits [11-15]. Although stability of a
certain polycyclic system is discussed in terms of contributions of
individual monocycles in both theories, the underlying models differ one
from another significantly: Pi-electron subsystems of (poly)cyclic
hydrocarbons are assumed to contain uniform carbon-carbon bonds as usual in
the theory of cyclic conjugation (cf. the molecular graph [15, 16]), whereas
conjugated circuits are defined as cycles consisting of strong (C=C) and
weak (C$-$C) bonds alternately and thereby refer mainly to separate Kekul%
\'{e} valence structures of the given hydrocarbon [15]. Again, the
above-mentioned definition allowed a fruitful analogy to be traced [17]
between the theory of conjugated circuits [abbreviated below as the CC
theory] and the perturbative treatment of pi-electron systems of conjugated
hydrocarbons based on a model of weakly-interacting initially-double (C=C)
bonds [18]. Moreover, this analogy along with the earlier-derived power
series for total energies of molecules [19] resulted into an original
perturbative approach [17, 20, 21] that exhibited an additional cognitive
and discriminative potential vs. that of the CC theory. The present
contribution contains a continuation of studies in this direction.

The CC theory of polycyclic conjugated hydrocarbons is currently under an
intensive developement [22-33] and its principal achievements are discussed
in the review [15]. This theory was especially successful for benzenoid
hydrocarbons having conjugated circuits of the 4n+2 series only as proven in
[23] [these circuits are usually designated by R$_{n}$ (n=1,2...), where R$%
_{1}$ coincides with a single Kekul\'{e} valence structure of benzene, R$%
_{2} $ embraces five C=C and five C$-$C bonds alternately, etc.]. So far as
non-benzenoid hydrocarbons, in general, and biphenylene-like systems, in
particular, are concerned, the results of the same theory proved to be
considerably less satisfactory [15, 33]. This especially refers to the
simplest version of the CC theory, wherein all Kekul\'{e} valence structures
of the given compound are assumed to be of the same "weight". Otherwise,
reliable criteria are required to discriminate between importances of
separate structures. Unfortunately, different criteria sometimes yield
contradictory conclusions [15, 34].

Biphenylene and its derivatives (e.g. [N]phenylenes) [35] evidently are of a
more involved constitution as compared to benzenoids. Indeed, these
non-benzenoid molecules contain four-atomic rings along with the six-atomic
ones and thereby conjugated circuits (CCs) of both R$_{n}$ ($4n+2$) and Q$%
_{n}$ ($4n$) series, the latter notation standing for circuits embracing
even numbers of C=C (and thereby of C$-$C) bonds.\ Thus, an interplay of
opposite factors is expected to determine the actual stabilities of
pi-electron systems in this case [10, 36] [Six- and four-atomic rings are
assumed to contribute to stabilization and to destabilization, respectively,
in accordance with the H\"{u}ckel [$4n+2$] rule]. Consequently, even the
extent and the nature of aromaticity of the parent biphenylene are still
under discussion [15, 36-38], to say nothing about its derivatives.
Dependence of the total pi-electron energy upon the Kekul\'{e} structure
count of phenylenes also were found to differ essentially from that of
benzenoids [39, 40]. In this context, the above-discussed difficulties of
the CC theory cause little surprise. To be able to foresee the most
efficient ways of improvement of the theory, however, a more specific
(preferably a deductive) accounting is highly desirable for its lower
success in the case of biphenylene-like hydrocarbons. To achieve this end,
just the above-mentioned perturbative approach seems to be helpful.

The total energy of a pi-electron system ($\mathcal{E}$) has been expressed
in this approach [17, 19-21] as a sum of steadily diminishing increments ($%
\mathcal{E}_{(k)}$) of various orders ($k$) with respect to the averaged
resonance parameter ($\gamma $) representing the weak (C$-$C) bonds.
Application of this approach to isolated CCs (R$_{1}$ and R$_{2}$) indicated
significant positive corrections of odd orders ($\mathcal{E}_{(3)}$ and $%
\mathcal{E}_{(5)},$\ respectively) to arise in the relevant power series
that are responsible for the excessive stability of these monocycles [Note
that a negative energy unit was used]. Moreover, parallelism has been
established between separate terms of the series for total energies of
individual Kekul\'{e} valence structures of benzenoid hydrocarbons and the
circuits R$_{n}$ $\ (n=1,2,.$.) present there. Hence, the perturbative
approach of Refs.[17, 19-21] proved to be a certain deductive analogue of
the CC theory that is likely to make possible an independent verification of
its principal assumptions. In this context, an important question naturally
arises whether the standard circuits R$_{n}$\ and Q$_{n}$\ ($n=1,2...$) are
the only important fragments (substructures) determining relative
stabilities of individual Kekul\'{e} valence structures of hydrocarbons
concerned. This point also is under focus of the present study.

In summary, our aim consists in application of the approach of Refs. [17,
19-21] to individual Kekul\'{e} valence structures of biphenylene-like
hydrocarbons, as well as in a deductive revealing the principal fragments
(substructures) determining their total energies and thereby relative
stabilities.

The scheme of the paper is as follows: We start with an overview of the
principal formulae of the perturbative approach (Sect. 2). Thereupon, we
apply these formulae to individual Kekul\'{e} valence structures of
biphenylene and of some related hydrocarbons (Sect. 3). The expected
decisive substructures are considered separately in Section 4$.$ Section 5
addresses the simplest combinations of these substructures, whilst the final
Section (6) contains the conclusions.

\begin{center}
\textbf{2. The principal formulae of the perturbative approach }
\end{center}

As already mentioned, the approach of Refs.[17, 19-21] addressed pi-electron
systems of conjugated hydrocarbons containing two types of uniform bonds,
namely strong (C=C) and weak (C$-$C) ones. The power series for total
energies of these systems has been derived and analyzed previously. Thus, we
will confine ourselves here to a brief overview of the principal definitions
and formulae.

Let us start with the basis set $\{\varphi \}$ underlying the series
concerned and consisting of bond orbitals (BOs) of strong (C=C) bonds. The
bonding BO (BBO)\ and the antibonding BO (ABO) of a certain C=C bond are
correspondingly defined as a normalized sum and difference of pairs of $%
2p_{z}$\ AOs of carbon atoms involved in the given bond. These BOs will be
denoted by $\varphi _{(+)i}$ and $\varphi _{(-)i}$, respectively, where the
subscript $i$ refers to the Ith C=C bond.\ Let the number of C=C bonds to
coincide with $N$. The set of BOs $\{\varphi \}$ then accordingly contains $%
2N$ basis functions.

The power series for total energies of the above-specified pi-electron
systems ($\mathcal{E}$) takes the form%
\begin{equation}
\mathcal{E}=\mathop{\displaystyle \sum }\limits_{k=0}^{\infty }\mathcal{E}%
_{(k)}
\end{equation}%
and contains the following starting members%
\begin{equation}
\mathcal{E}_{(0)}=2N,\quad \mathcal{E}_{(1)}=0,\quad \mathcal{E}_{(2)}=4Tr(%
\mathbf{G}_{(1)}\mathbf{G}_{(1)}^{+}),\quad \mathcal{E}_{(3)}=4Tr(\mathbf{G}%
_{(2)}\mathbf{G}_{(1)}^{+}),
\end{equation}%
where $Tr$ stands here and below for the $Trace$ of the whole matrix product
within parentheses, the superscript + designates a transposed
(Hermitian-conjugate) matrix, and $\mathbf{G}_{(1)}$\ and $\mathbf{G}_{(2)}$%
\ are the principal matrices of our expansion of the first and second
orders, respectively [The order is defined with respect to parameter $\gamma 
$\ as demonstrated below]. Let $\mathbf{S,Q}$ and $\mathbf{R}$ stand for
matrices, individual elements of which represent distinct types of
interactions (resonance parameters) between BOs, viz. \ 
\begin{eqnarray}
\mathbf{S}_{ij} &=&<\varphi _{(+)i}\mid \widehat{H}\mid \varphi
_{(+)j}>,\qquad \mathbf{R}_{il}=<\varphi _{(+)i}\mid \widehat{H}\mid \varphi
_{(-)l}>,  \notag \\
\mathbf{Q}_{lm} &=&<\varphi _{(-)l}\mid \widehat{H}\mid \varphi _{(-)m}>,
\end{eqnarray}%
where the orbitals concerned are shown inside the bra- and ket-vectors. The
principal matrices $\mathbf{G}_{(1)}$\ and $\mathbf{G}_{(2)}$\ are then
expressible via matrices $\mathbf{S,Q}$ and $\mathbf{R,}$ viz.\ 
\begin{equation}
\mathbf{G}_{(1)}=-\frac{1}{2}\mathbf{R,\quad G}_{(2)}=\frac{1}{4}(\mathbf{SR}%
-\mathbf{RQ})=\frac{1}{4}(\mathbf{SR}+\mathbf{RS)},
\end{equation}%
where the last relation is based on the equality $\mathbf{S=-Q}$\ valid for
even alternant \ hydrocarbons [1, 2] [Kekul\'{e} valence structures to be
studied belong to].

In contrast to the simple and unique formulae for $\mathcal{E}_{(2)}$ and $%
\mathcal{E}_{(3)}$ shown in Eq.(2), alternative expressions are possible for
the relevant fourth order term $\mathcal{E}_{(4)}$. Let us dwell here on the
formula for $\mathcal{E}_{(4)}$\ in the form of a sum of two components of
opposite signs [20], viz. of the positive component ($\mathcal{E}%
_{(4)}^{(+)} $) and of the negative one ($\mathcal{E}_{(4)}^{(-)}$) defined
as follows

\begin{equation}
\mathcal{E}_{(4)}^{(+)}=4Tr(\mathbf{G}_{(2)}\mathbf{G}_{(2)}^{+})>0,\quad 
\mathcal{E}_{(4)}^{(-)}=-4Tr(\mathbf{G}_{(1)}\mathbf{G}_{(1)}^{+}\mathbf{G}%
_{(1)}\mathbf{G}_{(1)}^{+})<0.
\end{equation}

To make the application of Eqs.(1)-(5) to pi-electron systems of Kekul\'{e}
valence structures more convenient, direct expressions are desirable for
matrices $\mathbf{S,Q}$ and $\mathbf{R}$\ (and thereby for $\mathbf{G}_{(1)}$%
\ and $\mathbf{G}_{(2)}$) in terms of submatrices (blocks) of the relevant
initial Hamiltonian matrix represented in the basis of 2p$_{z}$ AOs of
carbon atoms $\{\chi \}$ as usual. Let us dwell now just on these
expressions.

Let the $2p_{z}$\ AOs be characterized by uniform Coulomb parameters ($%
\alpha $). Moreover, uniform resonance parameters ($\beta $) are supposed to
correspond to any C=C bond in the basis $\{\chi \}$. Let us also accept the
usual equalities $\alpha =0$\ and $\beta =1,$\ the latter implying a
negative energy unit to be actually chosen. The total 2$N$-dimensional basis
set $\{\chi \}$\ of an even alternant hydrocarbon\ is known to be divisible
into two $N$-dimensional subsets $\{\chi _{{}}^{\ast }\}$\ and $\{\chi
^{\circ }\}$\ so that pairs of AOs belonging to any chemical bond (either
C=C or C$-$C) find themselves in different subsets [1, 2, 41-43]. This
implies non-zero resonance parameters representing chemical bonds to take
place in the off-diagonal (intersubset) positions of the initial Hamiltonian
matrix ($\mathbf{H}$) [41-43]. Let these blocks to be denoted by $\mathbf{H}%
_{\ast \circ }.$ Given that pairs of AOs belonging to the same (say Ith) C=C
bond acquire coupled numbers $i$ and $N+i$ in addition, resonance parameters
of these strong bonds (coinciding with our energy unit $\beta $) take the
diagonal positions of the blocks $\mathbf{H}_{\ast \circ }$\ and thereby
compose a unit matrix ($\mathbf{I}$). We then obtain that%
\begin{equation}
\mathbf{H}_{\ast \circ }=\mathbf{I}+\gamma \mathbf{B,}
\end{equation}%
where the matrix $\mathbf{B}$\ contains unit elements in the positions
referring to C$-$C bonds and zero elements elsewhere. Meanwhile, $\gamma $\
represents the averaged resonance parameter of C$-$C bonds that is supposed
to take a small value in our energy units. The above-introduced matrices $%
\mathbf{S,Q}$ and $\mathbf{R}$\ are then proportional to the symmetric
(Hermitian) and skew-symmetric (skew-Hermitian) parts of the matrix $\mathbf{%
B,}$\ respectively, viz. 
\begin{equation}
\mathbf{S}=-\mathbf{Q}=\frac{\gamma }{2}(\mathbf{B}+\mathbf{B}^{+}),\qquad 
\mathbf{R}=\frac{\gamma }{2}(\mathbf{B}^{+}-\mathbf{B}).
\end{equation}%
The matrix $\mathbf{B}$\ is easily constructable for any Kekul\'{e} valence
structure. Expressions of Eq.(7) then yield matrices of interbond resonance
parameters ($\mathbf{S,Q}$ and $\mathbf{R}$), whilst Eq.(4) provides us with
the relevant principal matrices $\mathbf{G}_{(1)}$\ and $\mathbf{G}_{(2)}$\
to be subsequently substituted into Eqs. (1), (2) and (5) to derive the
total energy. The first and second orders of matrices $\mathbf{G}_{(1)}$\
and $\mathbf{G}_{(2)}$\ with respect to parameter $\gamma $\ easily follow
from Eqs. (4) and (7). Moreover, the skew-symmetric (skew-Hermitian) nature
of these matrices may be proven [43] that ensures vanishing diagonal
elements $\mathbf{G}_{(1)ii}$ and $\mathbf{G}_{(2)ii}$ for any $i$.

\ Let us turn now to interpretation of the above formulae and start with
elements of matrices $\mathbf{G}_{(1)}$\ and $\mathbf{G}_{(2)}\ $defined by
Eq.(4). The first order element ($\mathbf{G}_{(1)il}$) is proportional to
the relevant resonance parameter ($\mathbf{R}_{il}$) and represents the
direct (through-space) interaction between the BBO $\varphi _{(+)i}$ and the
ABO $\varphi _{(-)l}$. Let pairs of C=C bonds connected by a C$-$C bond be
regarded as first-neighboring. It is then evident that non-zero elements ($%
\mathbf{G}_{(1)il}\neq 0$ and $\mathbf{G}_{(1)li}\neq 0$)\ generally
correspond to first-neighboring pairs of C=C bonds and/or to individual C$-$%
C bonds [This does not imply, however, that any C$-$C bond necessarily is
represented by non-zero elements $\mathbf{G}_{(1)il}$ and $\mathbf{G}%
_{(1)li} $ \ as the results of Sect. 3 show]. Further, the second order
elements $\mathbf{G}_{(2)il}$ are accordingly interpretable as indirect
(through-bond) interactions of the same BOs. Indeed, from Eq.(4) we obtain%
\begin{equation}
\mathbf{G}_{(2)il}=\frac{1}{4}\left[ \mathop{\displaystyle \sum }%
\limits_{(+)j}\mathbf{S}_{ij}\mathbf{R}_{jl}-\mathop{\displaystyle \sum }%
\limits_{(-)m}\mathbf{R}_{im}\mathbf{Q}_{ml}\right] ,
\end{equation}%
where sums over $(+)j$ and over $(-)m$ correspondingly embrace BBOs ($%
\varphi _{(+)j}$) and ABOs ($\varphi _{(-)m}$). It is seen that BOs of other
bonds play the role of mediators in the second order interaction between
orbitals $\varphi _{(+)i}$ and $\varphi _{(-)l}$ [Note that elements $%
\mathbf{S}_{ii}\mathbf{,Q}_{ll}$ and $\mathbf{R}_{ii}$ vanish]. Moreover,
the orbitals $\varphi _{(+)j}$ and $\varphi _{(-)m}$ should overlap directly
both with $\varphi _{(+)i}$ and with $\varphi _{(-)l}$ to be efficient
mediators. That is why non-zero indirect interactions correspond to pairs of
second-neighboring C=C bonds possessing a common first neighbor.

Let us dwell now on members of the power series for total energies. As is
seen from the first relation of Eq.(2), the zero order member $\mathcal{%
\varepsilon }_{(0)}$ coincides with the total energy of $N$ isolated C=C
bonds in accordance with the expectation. The subsequent (second order) term 
$\mathcal{\varepsilon }_{(2)},$\ in turn, depends on the total number of
non-zero direct interactions ($\mathbf{G}_{(1)il}$) between BBOs and ABOs.
In the usual case of two-to-one correspondence between significant elements $%
\mathbf{G}_{(1)il}$\ ($\mathbf{G}_{(1)li}$) and C$-$C bonds [17, 21], the
second order energy $\mathcal{\varepsilon }_{(2)}$\ complies with the
relation $\mathcal{\varepsilon }_{(2)}=N^{\prime }\gamma ^{2}/2$ [e.g. $%
\mathcal{\varepsilon }_{(2)}(R_{1})=3\gamma ^{2}/2$ [17]], where $\gamma
^{2}/2$\ is the contribution of a single C$-$C bond and $N^{\prime }$\ here
and below stands for the number of these bonds [Note that numbers $N^{\prime
}$\ and $N$\ generally do not coincide one with another]. The \textit{a
priori} positive sign of the second order energy $\mathcal{\varepsilon }%
_{(2)}$\ and thereby its stabilizing nature also deserves adding here.

The third order member of the same series ($\mathcal{\varepsilon }_{(3)}$)
is determined by products $\mathbf{G}_{(2)il}\mathbf{G}_{(1)il}$ as Eq.(2)
shows. This implies this member to take a non-zero value if in the given
system there is at least a single pair of BOs ($\varphi _{(+)i}$ and $%
\varphi _{(-)l}$) that interact both directly and indirectly by means of a
single mediator. Since the latter necessarily belongs to a third (say, the
Mth) C=C bond (that coincides neither with the Ith bond nor with the Lth
one), we arrive at a condition of presence of at least a single triplet of
C=C bonds, the BOs of all pairs of which interact (overlap) directly, i.e.
of a triplet I, L, M, wherein all three pairs of C=C bonds are
first-neighboring. In the case of Kekul\'{e} valence structures of
benzenoids, any circuit R$_{1}$ evidently offers a required triplet.
Consequently, the relevant third order energies proved to be additive
quantities with respect to transferable increments of individual CCs R$_{1},$
each of these increments coinciding with $\mathcal{\varepsilon }%
_{(3)}(R_{1})=3\gamma ^{3}/4$ [17]$.$\ Given that $K$ stands for the total
number of the circuits R$_{1}$ in the given Kekul\'{e} valence structure,
the correction concerned meets the relation $\mathcal{\varepsilon }%
_{(3)}=3\gamma ^{3}K/4.$ Conditions determining the sign of the third order
energy also easily follow from the last formula of Eq.(2). Indeed, a
particular product $\mathbf{G}_{(2)il}$\ $\mathbf{G}_{(1)il}$\ yields a
positive (negative) contribution to the total correction $\mathcal{%
\varepsilon }_{(3)},$\ if the participating interbond interactions (i.e.
both $\mathbf{G}_{(2)il}$\ and $\mathbf{G}_{(1)il}$)\ are of the same
(opposite) sign(s). In particular, positive products $\mathbf{G}_{(2)il}$\ $%
\mathbf{G}_{(1)il}$\ only were shown to refer to any circuit R$_{1}$ in the
Kekul\'{e} valence structures of benzenoid hydrocarbons [17].

Finally, the fourth order correction $\mathcal{\varepsilon }_{(4)}$\ remains
to be discussed. As is seen from Eq.(5), the positive component $\mathcal{%
\varepsilon }_{(4)}^{(+)}$\ (that is stabilizing in our energy units) is
determined by squares of indirect interbond interactions $\mathbf{G}%
_{(2)il}. $\ Meanwhile, the relevant destabilizing component $\mathcal{%
\varepsilon }_{(4)}^{(-)}$\ is proportional to the sum of squares of
elements of the matrix $\mathbf{G}_{(1)}\mathbf{G}_{(1)}^{+}$. The latter
elements are interpretable as indirect interactions between BBOs ($\varphi
_{(+)i}$ and $\varphi _{(+)j}$) via ABOs playing the role of mediators. In
this connection, elements $\mathbf{G}_{(2)il}$\ and $(\mathbf{G}_{(1)}%
\mathbf{G}_{(1)}^{+})_{ij}$\ may be correspondingly referred to as
stabilizing and destabilizing indirect interactions. Accordingly, both the
absolute value and the sign of the correction $\mathcal{\varepsilon }_{(4)}$%
\ is determined by the balance between the above-specified two types of
interactions. It is no surprise in this connection that the overall
relations between the fourth order energies $\mathcal{\varepsilon }_{(4)}$\
and the presence of individual CCs proved to be much more involved as
compared to the above-discussed direct relation between $\mathcal{%
\varepsilon }_{(3)}$\ and the number of the circuits R$_{1}$ ($K$). For
isolated circuits R$_{n}$ (n=1,2,..), however, two simple rules have been
established [17]: First, all these circuits are characterized by positive
corrections $\mathcal{\varepsilon }_{(4)}$, i.e. $\mathcal{\varepsilon }%
_{(4)}(R_{n})>0$\ for any $n$. Second, the corrections concerned are equal
to $2\gamma ^{4}N/64$, where $2\gamma ^{4}/64$\ is the contribution of a
single C=C (or C$-$C) bond. This relation may be exemplified by the fourth
order energy of an isolated circuit R$_{1}$, viz. $\mathcal{\varepsilon }%
_{(4)}(R_{1})=6\gamma ^{4}/64$\ [Note that $\gamma ^{4}/64$\ has been chosen
as a convenient "subsidiary" unit for fourth order energies [17]].

\begin{center}
\textbf{3. Total energies of Kekul\'{e} valence structures of biphenylene
and related hydrocarbons}
\end{center}

Let us start with biphenylene itself (I). The four symmetry-non-equivalent
Kekul\'{e} valence structures of this hydrocarbon (I/I, I/II, I/III and
I/IV) are shown in Fig.1 along with their compositions in terms of standard
CCs R$_{n}$ and Q$_{n}$ ($n=1,2...$). Since these CCs correspondingly
embrace odd and even numbers of C=C bonds, these are conveniently referred
to below as odd- and even-membered CCs. Moreover, all Kekul\'{e} valence
structures of Fig. 1 contain six C=C bonds ($N=6$) and eight C$-$C bonds ($%
N\prime =8$). [It deserves attention that $N\neq N\prime $ ].

\begin{figure}[!ht]
	\begin{center}
	\vspace*{5 pt}
	\includegraphics[width=14cm]{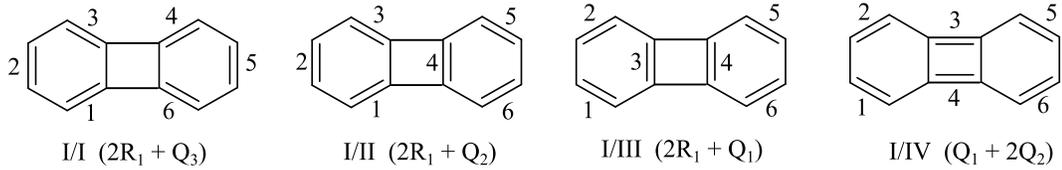}
  \caption{Symmetry-non-equivalent Kekul\'{e} valence structures of biphenylene
(I) along with their compositions in terms of standard conjugated circuits R$%
_{n}$ and Q$_{n},n=1,2,...$ . Numbers of C=C bonds also are shown.}
	\label{fig1}
	\end{center}
\end{figure}	

Zero order energies of structures I/I-I/IV coincide with 12 in our negative
energy units (see Eq.(2)). The subsequent non-zero energy corrections are as
follows%
\begin{eqnarray}
\mathcal{\varepsilon }_{(2)}(I/I) &=&\mathcal{\varepsilon }%
_{(2)}(I/II)=4\gamma ^{2},\quad \mathcal{\varepsilon }_{(2)}(I/III)=\mathcal{%
\varepsilon }_{(2)}(I/IV)=3\gamma ^{2}, \\
\mathcal{\varepsilon }_{(3)}(I/I) &=&\mathcal{\varepsilon }_{(3)}(I/III)=%
\frac{6\gamma ^{3}}{4},\quad \mathcal{\varepsilon }_{(3)}(I/II)=\frac{%
5\gamma ^{3}}{4},\quad \mathcal{\varepsilon }_{(3)}(I/IV)=0, \\
\mathcal{\varepsilon }_{(4)}(I/I){}\! &=&\!\!0,\mathcal{\varepsilon }%
_{(4)}(I/II)=-\frac{28\gamma ^{4}}{64},\mathcal{\varepsilon }_{(4)}(I/III)=%
\frac{44\gamma ^{4}}{64},\mathcal{\varepsilon }_{(4)}(I/IV)=-\frac{20\gamma
^{4}}{64}
\end{eqnarray}%
It is seen that second order energies of separate Kekul\'{e} valence
structures of biphenylene are not uniform in spite of the same number of C$-$%
C bonds ($N\prime $). Moreover, the usual relation $\mathcal{\varepsilon }%
_{(2)}=N^{\prime }\gamma ^{2}/2$ (Sect. 2) is met by the first two
corrections (viz. $\mathcal{\varepsilon }_{(2)}(I/I)$ and $\mathcal{%
\varepsilon }_{(2)}(I/II)$) but not by the remaining ones ($\mathcal{%
\varepsilon }_{(2)}(I/III)$ and $\mathcal{\varepsilon }_{(2)}(I/IV)$) that
are lower by $\gamma ^{2}$\ vs. the former. This implies a certain
exceptional second order destabilization to take place in the last two Kekul%
\'{e} valence structures of biphenylene. Since the latter contain a circuit Q%
$_{1}$ in contrast to structures I/I and I/II, the destabilization concerned
may be assumed to be related to the presence of just these simplest
even-membered CCs.

Let us comment now the remaining relations of Eq.(10). The corrections $%
\mathcal{\varepsilon }_{(3)}(I/I),$ $\mathcal{\varepsilon }_{(3)}(I/III)$
and $\mathcal{\varepsilon }_{(3)}(I/IV)$\ now\ comply with the usual
relation between the third order energy $(\mathcal{\varepsilon }_{(3)})$\
and the total number $K$\ of the simplest odd-membered circuits R$_{1}$
(Sect. 2), where $K$ coincides with 2, 2 and 0, respectively. Meanwhile, the
second structure of biphenylene (I/II) is an exception in this respect.
Indeed, the correction $\mathcal{\varepsilon }_{(3)}(I/II)$\ is lower vs.
that following from the relation $\mathcal{\varepsilon }_{(3)}=3\gamma
^{3}K/4$\ for $K=2$. This fact gives us a hint that some additional
structural factor(s) and/or substructure(s) exert a destabilizing influence
upon the third order energy $\mathcal{\varepsilon }_{(3)}(I/II).$\ An
important point here is that the even-membered conjugated circuit Q$_{2}$ of
the structure I/II hardly is able to play the above-anticipated role, as it
does not offer a necessary triplet of first-neighboring C=C bonds (Sect. 2).
Finally, the lowest (negative) fourth order energies are peculiar to the Q$%
_{2}$-containing structures I/II and I/IV. Thus, a certain fourth order
destabilization may be assumed to take place due to the presence of a
four-membered circuit Q$_{2}$. Meanwhile, the highest (positive) correction $%
\mathcal{\varepsilon }_{(4)}(I/III)$ refers to the structure I/III
containing the simplest CCs R$_{1}$ and Q$_{1}$ only.

In summary, the above results suggest the following order of relative
stability for the Kekul\'{e} valence structures of biphenylene: $%
I/I>I/II>I/III>I/IV$ [Note that sums of second and third order energies
actually are sufficient to discriminate between the structures concerned].
Applications of other criteria also yield the same conclusion, e.g.
comparison of contributions of separate Kekul\'{e} valence structures to the
resonance energy of biphenylene [15], as well as invoking of additional
concepts, such as parity [44], innate degrees of freedom [45] and Kekul\'{e}
index [46, 47] of a Kekul\'{e} valence structure. Finally, a similar order
of stability has been established recently on the basis of a modified
Hess-Schaad group additivity scheme [36].

\begin{figure}[!ht]
	\begin{center}
	\vspace*{5 pt}
	\includegraphics[width=14cm]{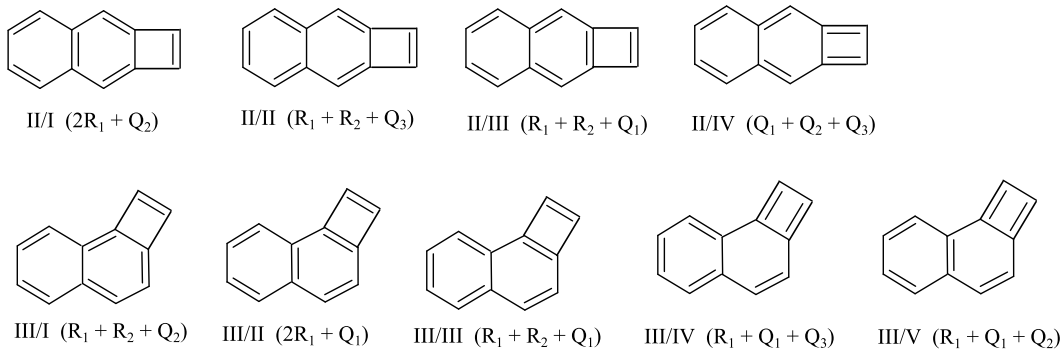}
  \caption{Kekul\'{e} valence structures of naphthocyclobutenes (II) and (III)
along with their compositions in terms of standard conjugated circuits R$%
_{n} $ and Q$_{n},n=1,2,.$.}
	\label{fig2}
	\end{center}
\end{figure}	

\ Let us turn now to the naphthocyclobutene (II) and consider its Kekul\'{e}
valence structures II/I-II/IV (Fig. 2). Zero order energies of these
structures also are equal to 12 so that their stabilities are directly
comparable to those of the former structures of biphenylene. Further, the Q$%
_{1}$-containing structures II/III and II/IV are characterized by an
exceptionally low second order energy in this case as well, viz. 
\begin{equation}
\mathcal{\varepsilon }_{(2)}(II/I)=\mathcal{\varepsilon }_{(2)}(II/II)=4%
\gamma ^{2},\quad \mathcal{\varepsilon }_{(2)}(II/III)=\mathcal{\varepsilon }%
_{(2)}(II/IV)=3\gamma ^{2}
\end{equation}%
and the destabilization effect coincides with $\gamma ^{2}$\ as previously.\
For the relevant third order corrections, we obtain%
\begin{equation}
\mathcal{\varepsilon }_{(3)}(II/I)=\frac{5\gamma ^{3}}{4},\quad \mathcal{%
\varepsilon }_{(3)}(II/II)=\frac{2\gamma ^{3}}{4},\quad \mathcal{\varepsilon 
}_{(3)}(II/III)=\frac{3\gamma ^{3}}{4},\quad \mathcal{\varepsilon }%
_{(3)}(II/IV)=0.
\end{equation}%
It is seen that third order energies $\mathcal{\varepsilon }_{(3)}(II/I)$
and $\mathcal{\varepsilon }_{(3)}(II/II)$\ are now exceptionally lowered vs.
the anticipated values for $K=2$ and $K=1,$ respectively. Nevertheless, the
correction $\mathcal{\varepsilon }_{(3)}(II/I)$ of the 2R$_{1}$-containing
structure II/I exceeds that of the R$_{1}$-containing one ($\mathcal{%
\varepsilon }_{(3)}(II/II)$) and, consequently, the resulting order of
stability is expected to be as follows:\ $II/I>II/II>II/III>II/IV$. This
order cincides with that following from Kekul\'{e} indices [46]. Finally,
the relevant fourth order energies take the form%
\begin{eqnarray}
\mathcal{\varepsilon }_{(4)}(II/I) &=&-\frac{28\gamma ^{4}}{64},\quad 
\mathcal{\varepsilon }_{(4)}(II/II)=\frac{16\gamma ^{4}}{64},\quad \mathcal{%
\varepsilon }_{(4)}(II/III)=\frac{32\gamma ^{4}}{64},  \notag \\
\quad \mathcal{\varepsilon }_{(4)}(II/IV) &=&-\frac{4\gamma ^{4}}{64},
\end{eqnarray}%
wherein the lowest (negative) values refer to Q$_{2}$-containing structures
(II/I and II/IV) in this case too. Meanwhile, the fourth order energy of the
structure II/III takes the highest (positive) value. In this respect,
similarity between Kekul\'{e} valence structures I/III and II/III is
evident. Since the latter contain the simplest circuits R$_{1}$ and Q$_{1},$
the neighboring pairs of which possess a single common C=C bond, the high
(positive) values of the fourth order energies $\mathcal{\varepsilon }%
_{(4)}(I/III)$ and $\mathcal{\varepsilon }_{(4)}(II/III)$\ may be expected
to originate just from this common aspect of constitution. Furthermore, the
structures I/II and II/I offer us an example of uniform corrections
referring to distinct hydrocarbons, namely these structures prove to be
isoenergetic to within fourth order terms inclusive. Besides, the relevant
compositions in terms of standard CCs (2R$_{1}$+Q$_{2}$) also are uniform.

Our next example coincides with the bent isomer of naphthocyclobutene (III)
characterized by five Kekul\'{e} valence structures III/I-III/V (Fig. 2),
where $N=6$ and $N\prime =8$. The number of Q$_{1}$-containing structures
also is higher here, and this fact is reflected in the relevant second order
energies, viz. 
\begin{equation}
\mathcal{\varepsilon }_{(2)}(III/I)=4\gamma ^{2},\mathcal{\varepsilon }%
_{(2)}(III/II)=\mathcal{\varepsilon }_{(2)}(III/III)=\mathcal{\varepsilon }%
_{(2)}(III/IV)=\mathcal{\varepsilon }_{(2)}(III/V)=3\gamma ^{2}.
\end{equation}%
It is seen that the only Q$_{1}$-free structure (III/I) is desribed by a
"normal" energy correction in this case. The relevant third order energies
are as follows%
\begin{eqnarray}
\mathcal{\varepsilon }_{(3)}(III/I) &=&\frac{2\gamma ^{3}}{4},\qquad 
\mathcal{\varepsilon }_{(3)}(III/II)=\frac{6\gamma ^{3}}{4},  \notag \\
\mathcal{\varepsilon }_{(3)}(III/III) &=&\mathcal{\varepsilon }%
_{(3)}(III/IV)=\mathcal{\varepsilon }_{(3)}(III/V)=\frac{3\gamma ^{3}}{4}
\end{eqnarray}%
and the correction $\mathcal{\varepsilon }_{(3)}(III/I)$\ now takes a
lowered value as compared to that following from the relation $\mathcal{%
\varepsilon }_{(3)}=3\gamma ^{3}K/4$\ for $K=1$. Moreover, sums of energy
increments of second and third orders are insufficient to discriminate
between all Kekul\'{e} valence structures of the bent isomer III, in
contrast to the former hydrocarbons I and II. Fortunately, the relevant
fourth order corrections take distinct values, viz.%
\begin{eqnarray}
\mathcal{\varepsilon }_{(4)}(III/I) &=&-\frac{20\gamma ^{4}}{64},\quad 
\mathcal{\varepsilon }_{(4)}(III/II)=\frac{28\gamma ^{4}}{64},\quad \mathcal{%
\varepsilon }_{(4)}(III/III)=\frac{32\gamma ^{4}}{64},\quad   \notag \\
\mathcal{\varepsilon }_{(4)}(III/IV) &=&\frac{20\gamma ^{4}}{64},\quad 
\mathcal{\varepsilon }_{(4)}(III/V)=-\frac{8\gamma ^{4}}{64}
\end{eqnarray}%
and thereby allow us to discriminate them. The result of Eq.(17) is even
more surprising if we recall that the extended CCs neither R$_{2}$\ nor Q$%
_{3}$ may be entirely embraced by corrections of the fourth order (the
highest number of the embraced C=C bonds coincides with the order $k$ of the
relevant correction $\mathcal{\varepsilon }_{(k)}$). Equation (17) shows in
addition that the lowest (negative) values of the fourth order energy refer
to Q$_{2}$-containing structures III/I and III/V as previously. Meanwhile,
the remaining (R$_{1}$ and Q$_{1}$ -containing) structures III/II , III/III
and III/IV are characterized by positive fourth order energies. The overall
order of relative stabilities is then as follows: $%
III/I>III/II>III/III>III/IV>III/V$. An analogous conclusion has been drawn
also on the basis of Kekul\'{e} indices [46] and connectivities of the
relevant submolecules [48].

\begin{figure}[!ht]
	\begin{center}
	\vspace*{5 pt}
	\includegraphics[width=14cm]{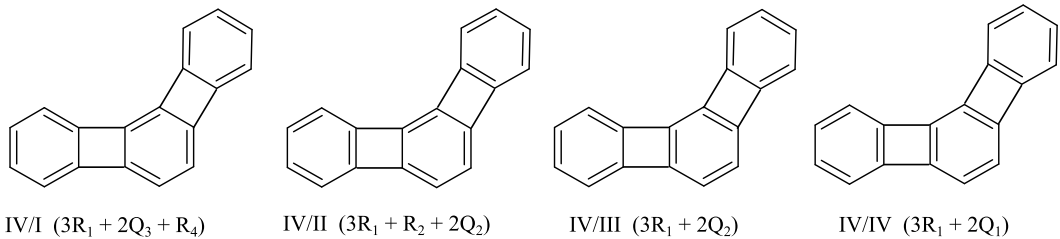}
  \caption{Selected Kekul\'{e} valence structures of bent [3]-phenylene along
with their compositions in terms of standard conjugated circuits R$_{n}$ and
Q$_{n},n=1,2,...$.}
	\label{fig3}
	\end{center}
\end{figure}	

Let us consider finally four selected Kekul\'{e} valence structures of bent
[3]phenylene IV/I-IV/IV (Fig. 3), where $N=9$ and $N\prime =13$. The common
zero order energy of these structures accordingly equals to 18, whilst the
second order corrections are as follows 
\begin{equation}
\mathcal{\varepsilon }_{(2)}(IV/I)=\mathcal{\varepsilon }_{(2)}(IV/II)=%
\mathcal{\varepsilon }_{(2)}(IV/III)=\frac{13}{2}\gamma ^{2},\quad \mathcal{%
\varepsilon }_{(2)}(IV/IV)=\frac{9}{2}\gamma ^{2}.
\end{equation}%
It is seen that the only Q$_{1}$-containing structure IV/IV is now
characterized by an exceptionally lowered third order energy in accordance
with the above-established trends. It is also noteworthy that the difference
between corrections shown in Eq.(18) coincides with $2\gamma ^{2}$\ instead
of the former distinction $\gamma ^{2}.$\ This fact is in line with two
conjugated circuits Q$_{1}$ contained in the structure IV/IV and thereby it
supports the above assumption about the relation of the difference concerned
to a destabilizing increment of just this circuit. The relevant third order
energies, in turn, equal to%
\begin{equation}
\mathcal{\varepsilon }_{(3)}(IV/I)=\mathcal{\varepsilon }_{(3)}(IV/IV)=\frac{%
9\gamma ^{3}}{4},\quad \mathcal{\varepsilon }_{(3)}(IV/II)=\mathcal{%
\varepsilon }_{(3)}(IV/III)=\frac{7\gamma ^{3}}{4}
\end{equation}%
and take non-uniform values in spite of the same number of the circuits R$%
_{1}$ ($K=3$) in the structures concerned. Moreover, two structures (viz.
IV/II and II/III) are now characterized by lowered third order energies ($%
7\gamma ^{3}/4$) that do not follow from the standard relation for $K=3$
and, consequently, just the above-mentioned two exceptional structures
cannot be discriminated before taking into account the relevant fourth order
corrections. These are as follows%
\begin{eqnarray}
\mathcal{\varepsilon }_{(4)}(IV/I) &=&-\frac{2\gamma ^{4}}{64},\quad 
\mathcal{\varepsilon }_{(4)}(IV/II)=-\frac{58\gamma ^{4}}{64},\quad \mathcal{%
\varepsilon }_{(4)}(IV/III)=-\frac{62\gamma ^{4}}{64},\quad  \notag \\
\mathcal{\varepsilon }_{(4)}(IV/IV) &=&\frac{82\gamma ^{4}}{64}
\end{eqnarray}%
and indicate the structure IV/II to be a little bit less destabilized (and
thereby more stable) as compared to IV/III. This result is in line with the
presence of an additional "stabilizing" circuit R$_{2}$ in the structure
IV/II.

The overall set of fourth order energies of Eq.(20) also deserves some
attention. Thus, the lowest (negative) values of these corrections (about $%
-60\gamma ^{4}/64$) refer to Q$_{2}$-containing structures (IV/II and
IV/III) as previously, whereas the highest (positive) value ($82\gamma
^{4}/64$) corresponds to the (3R$_{1}$+2Q$_{1})$-containing structure IV/IV.
Meanwhile, the fourth order energy of the remaining structure (IV/I) takes
an intermediate value. The latter, however, seems to be sufficiently low if
we recall that the structure IV/I contains no "destabilizing" circuits Q$_{2}
$. A more detailed discussion of this point is undertaken in Section 5.
Summarizing the results of Eqs.(18)-(20) we may then conclude the predicted
relative order of stabilities of the Kekul\'{e} valence structures under
discussion to be as follows: $IV/I>IV/II>IV/III>IV/IV$. This expectation is
in line with the actual contributions of the structures concerned to the
molecular resonance energy [15, 34].

On the whole, two principal assumptions (hypotheses) follow from the results
of this section: First, the even-membered conjugated circuits Q$_{1}$ and Q$%
_{2}$\ seem to contribute to the second and fourth order destabilization,
respectively, of the Kekul\'{e} valence structures concerned. Second, some
additional factors (substructures) are likely to participate in the
formation of third order energies of a part of Kekul\'{e} valence structures
of biphenylene-like hydrocarbons (along with standard conjugated circuits R$%
_{1}$) and thereby to be responsible for an extra third order
destabilization of these structures. Our next aim consists in verification
of these hypothesis and in revealing the above-anticipated additional
substructures.

\begin{center}
\textbf{4. Consideration of reference structures}
\end{center}

The present section is devoted to isolated substructures that are expected
to participate in the formation of the above-overviewed energy corrections.
The odd-membered CCs (R$_{n}$) have been studied previously in a detail [17]
and the relevant energy increments are overviewed in Section 2. Thus, let us
turn immediately to the even-membered CCs (Q$_{n},$ where $n=1,2...$).

\begin{center}
\textbf{4.1. Even-membered conjugated circuits Q}$_{n}$\textbf{\ }
\end{center}

Let us start with the simplest two-membered circuit Q$_{1}$ (Fig. 4), where $%
N=N\prime =2$ and $\mathcal{E}_{(0)}(Q_{1})=4$ (see Eq.(2)). The relevant
matrix $\mathbf{B(}Q_{1}\mathbf{)}$ is as follows%
\begin{equation}
\mathbf{B(}Q_{1}\mathbf{)=}\gamma \left\vert 
\begin{array}{cc}
0 & 1 \\ 
1 & 0%
\end{array}%
\right\vert ,
\end{equation}%
where unit elements represent resonance parameters of C$-$C bonds (viz. C$%
_{1}-$C$_{4}$ and C$_{2}-$C$_{3}$). The skew-symmetric part of this
symmetric matrix evidently vanishes. From Eqs. (4) and (7), we then obtain%
\begin{equation}
\mathbf{R(}Q_{1}\mathbf{)=G}_{(1)}\mathbf{(}Q_{1}\mathbf{)=G}_{(2)}\mathbf{(}%
Q_{1}\mathbf{)=G}_{(1)}\mathbf{G}_{(1)}^{+}\mathbf{(}Q_{1}\mathbf{)=0}
\end{equation}%
and 
\begin{equation}
\mathcal{E}_{(2)}(Q_{1})=\mathcal{E}_{(3)}(Q_{1})=\mathcal{E}_{(4)}(Q_{1})=0.
\end{equation}%

\begin{figure}[!ht]
	\begin{center}
	\vspace*{5 pt}
	\includegraphics[width=12cm]{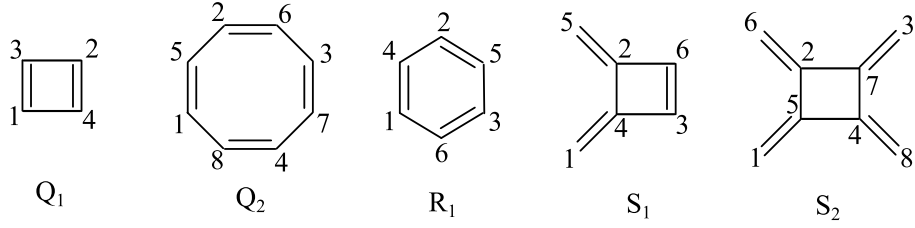}
  \caption{The simplest conjugated circuits of different series (R$_{n}$, Q$%
_{n} $ and S$_{n}).$ Numbers of carbon atoms and/or of their 2p$_{z}$ AOs
also are shown.}
	\label{fig4}
	\end{center}
\end{figure}	

If we recall here that the linear butadiene is characterized by a positive
second order energy equal to $\gamma ^{2}/2$ [17, 20], destabilization in
the circuit Q$_{1}$\ vs. its acyclic analogue may be concluded. This result
may be traced back to vanishing direct interactions between BBOs and ABOs of
the circuit Q$_{1}$ in contrast to the linear butadiene. Again, the zero
second order energy $\mathcal{E}_{(2)}(Q_{1})$\ seen from Eq.(23) seems to
be inconsistent with the rule about each C$-$C bond contributing $\gamma
^{2}/2$\ to the relevant total $\mathcal{E}_{(2)}$\ value (Section 2). To
clarify this point, let us represent the matrix $\mathbf{B(}Q_{1}\mathbf{)}$%
\ of\ \ Eq.(21) as a sum of two components%
\begin{equation}
\mathbf{B(}Q_{1}\mathbf{)=B}^{(1-4)}\mathbf{(}Q_{1}\mathbf{)+B}^{(2-3)}%
\mathbf{(}Q_{1}\mathbf{)=}\gamma \left\vert 
\begin{array}{cc}
0 & 1 \\ 
0 & 0%
\end{array}%
\right\vert +\gamma \left\vert 
\begin{array}{cc}
0 & 0 \\ 
1 & 0%
\end{array}%
\right\vert
\end{equation}%
containing resonance parameters of individual weak bonds C$_{1}-$C$_{4}$ and
C$_{2}-$C$_{3}$, as indicated by superscripts (1-4) and (2-3), respectively.
Moreover, the components $\mathbf{B}^{(1-4)}\mathbf{(}Q_{1}\mathbf{)}$\ and $%
\mathbf{B}^{(2-3)}\mathbf{(}Q_{1}\mathbf{)}$\ resemble the relevant matrix
of the linear butadiene [20]. An analogous partition of the matrix $\mathbf{G%
}_{(1)}\mathbf{(}Q_{1}\mathbf{)}$\ takes the form%
\begin{equation}
\mathbf{G}_{(1)}\mathbf{(}Q_{1}\mathbf{)=G}_{(1)}^{(1-4)}\mathbf{(}Q_{1}%
\mathbf{)+G}_{(1)}^{(2-3)}\mathbf{(}Q_{1}\mathbf{)=-}\frac{\gamma }{4}%
\left\vert 
\begin{array}{cc}
0 & -1 \\ 
1 & 0%
\end{array}%
\right\vert \mathbf{-}\frac{\mathbf{\gamma }}{4}\left\vert 
\begin{array}{cc}
0 & 1 \\ 
-1 & 0%
\end{array}%
\right\vert .
\end{equation}%
The second order energy $\mathcal{E}_{(2)}(Q_{1})$\ is then representable as
follows%
\begin{equation}
\mathcal{E}_{(2)}(Q_{1})=\mathcal{E}_{(2)}^{(1-4)}(Q_{1})+\mathcal{E}%
_{(2)}^{(2-3)}(Q_{1})+\mathcal{E}_{(2)}^{(cycl)}(Q_{1}),
\end{equation}%
where 
\begin{equation}
\mathcal{E}_{(2)}^{(1-4)}(Q_{1})=4Tr(\mathbf{G}_{(1)}^{(1-4)}\cdot \mathbf{G}%
_{(1)}^{(1-4)+}),\quad \mathcal{E}_{(2)}^{(2-3)}(Q_{1})=4Tr(\mathbf{G}%
_{(1)}^{(2-3)}\cdot \mathbf{G}_{(1)}^{(2-3)+})
\end{equation}%
are the contributions of individual C$-$C bonds. Meanwhile, the last term of
Eq.(26) takes the form%
\begin{equation}
\mathcal{E}_{(2)}^{(cycl)}(Q_{1})=4Tr(\mathbf{G}_{(1)}^{(1-4)}\cdot \mathbf{G%
}_{(1)}^{(2-3)+}+\mathbf{G}_{(1)}^{(2-3)}\cdot \mathbf{G}_{(1)}^{(1-4)+})%
\quad
\end{equation}%
and may be interpreted as the cyclization energy of the circuit Q$_{1}$.
[Besides, this term may be traced back to the so-called roundabout pathways
over BOs [49, 50]]. Substituting the expressions for $\mathbf{G}%
_{(1)}^{(1-4)}$\ and $\mathbf{G}_{(1)}^{(2-3)}$\ of Eq.(25) into Eqs.(27)
and (28) shows the energy components to take the following values%
\begin{equation}
\mathcal{E}_{(2)}^{(1-4)}(Q_{1})=\mathcal{E}_{(2)}^{(2-3)}(Q_{1})=\frac{%
\gamma ^{2}}{2},\quad \mathcal{E}_{(2)}^{(cycl)}(Q_{1})=-\gamma ^{2}.
\end{equation}%
so that each C$-$C bond contributes $\gamma ^{2}/2$\ to the second order
energy $\mathcal{E}_{(2)}(Q_{1})$\ in accordance with the above-discussed
rule. Again, the total increment of both C$-$C bonds coincides with the
absolute value of the (negative) cyclization energy $\mathcal{E}%
_{(2)}^{(cycl)}(Q_{1})$\ and, consequently, the correction $\mathcal{E}%
_{(2)}(Q_{1})$\ vanishes. Thus, the second order destabilization of the
circuit Q$_{1}$ due to cyclization is now even more evident.

The four-membered circuit Q$_{2}$ (Fig. 4) also may be studied similarly,
where $N=N\prime =4$ and $\mathcal{E}_{(0)}(Q_{2})=8$. Further, pairs of
neighboring C=C bonds build up linear butadiene-like fragments in the
circuit Q$_{2}$ in contrast to Q$_{1},$ and, consequently, the relevant BBOs
and ABOs interact directly. As a result, both the first order matrix $%
\mathbf{G}_{(1)}\mathbf{(}Q_{2}\mathbf{)}$\ and its derivative $\mathbf{G}%
_{(1)}\mathbf{G}_{(1)}^{+}\mathbf{(}Q_{2}\mathbf{)}$\ are non-zero matrices,
viz. 
\begin{equation}
\mathbf{G}_{(1)}\mathbf{(}Q_{2}\mathbf{)=-}\frac{\gamma }{4}\left\vert 
\begin{array}{cccc}
0 & 1 & 0 & -1 \\ 
-1 & 0 & 1 & 0 \\ 
0 & -1 & 0 & 1 \\ 
1 & 0 & -1 & 0%
\end{array}%
\right\vert ,\mathbf{G}_{(1)}\mathbf{G}_{(1)}^{+}\mathbf{(}Q_{2}\mathbf{)=}%
\frac{\gamma ^{2}}{16}\left\vert 
\begin{array}{cccc}
2 & 0 & -2 & 0 \\ 
0 & 2 & 0 & -2 \\ 
-2 & 0 & 2 & 0 \\ 
0 & -2 & 0 & 2%
\end{array}%
\right\vert
\end{equation}%
and yield the following energy increments 
\begin{equation}
\mathcal{E}_{(2)}(Q_{2})=2\gamma ^{2},\quad \mathcal{E}_{(4)}^{(-)}(Q_{2})=-%
\frac{32\gamma ^{4}}{64}.
\end{equation}%
Thus, the second order energy $\mathcal{E}_{(2)}(Q_{2})$\ coincides with the
four-fold increment of a single C$-$C bond ($\gamma ^{2}/2$) in accordance
with our rule.

In contrast to $\mathbf{G}_{(1)}\mathbf{(}Q_{2}\mathbf{)}$\ of Eq.(30), the
second order matrix $\mathbf{G}_{(2)}\mathbf{(}Q_{2}\mathbf{)}$\ vanishes
for the circuit Q$_{2}.$\ The underlying reason is that mediating effects of
the intervening C=C bonds cancel out one another when building up elements
of the matrix $\mathbf{G}_{(2)}\mathbf{(}Q_{2}\mathbf{)}$\ referring to BOs
of second-neighboring C=C bonds. For example, the mediating effects of BOs
of C$_{2}$=C$_{6}$ and C$_{4}$=C$_{8}$ bonds\ are of coinciding absolute
values and of opposite signs in the expression for the element $\mathbf{G}%
_{(2)13}\mathbf{(}Q_{2}\mathbf{).}$\ Consequently, both the third order
energy $\mathcal{E}_{(3)}(Q_{2})$\ and the fourth order stabilizing
component $\mathcal{E}_{(4)}^{(+)}(Q_{2})$\ also take zero values. The total
fourth order energy $\mathcal{E}_{(4)}(Q_{2})$ then coincides with $\mathcal{%
E}_{(4)}^{(-)}(Q_{2})$\ of Eq.(31) and is a negative quantity, i.e.%
\begin{equation}
\mathcal{E}_{(4)}(Q_{2})=-\frac{32\gamma ^{4}}{64}.
\end{equation}%
If we recall that the acyclic isomer of octatetraene is characterized by a
positive fourth order energy $2\gamma ^{4}/64$\ [20], the above result
implies a fourth order destabilization of the circuit Q$_{2}$. Partition of
the correction $\mathcal{E}_{(4)}(Q_{2})$\ into increments of cyclic and
acyclic origin (like that performed for $\mathcal{E}_{(2)}(Q_{1})$) shows
that the cyclization energy of the circuit Q$_{2}$ equals to $-40\gamma
^{4}/64$. Meawhile, the total increment of acyclic nature coincides with $%
8\gamma ^{4}/64,$\ i.e. with the four-fold contribution of a single C=C (or C%
$-$C) bond [as it was the case with the odd-membered circuits R$_{n}$ [17]].

Finally, the six-membered circuit Q$_{3}$ deserves some attention. Matrices
both $\mathbf{G}_{(1)}\mathbf{(}Q_{3}\mathbf{)}$\ and $\mathbf{G}_{(2)}%
\mathbf{(}Q_{3}\mathbf{)}$\ do not vanish in this case, and this result
causes no surprise. Non-zero elements, however, take distinct positions in
these matrices and, consequently, the third order energy $\mathcal{E}%
_{(3)}(Q_{3})$\ vanishes. Meanwhile, the corrections $\mathcal{E}%
_{(2)}(Q_{3})$\ and $\mathcal{E}_{(4)}(Q_{3})$\ are as follows%
\begin{equation}
\mathcal{E}_{(2)}(Q_{3})=3\gamma ^{2},\quad \mathcal{E}_{(4)}(Q_{3})=\frac{%
12\gamma ^{4}}{64}
\end{equation}%
and coincide with respective six-fold increments of an individual C=C (or C$%
- $C) bonds. Hence, the destabilizing effect of cyclization (if any) seems
to manifest itself within corrections of higher orders in this case.

Therefore, additional destabilizing contributions arise within corrections $%
\mathcal{E}_{(2)}$\ and $\mathcal{E}_{(4)}$\ of the circuits Q$_{1}$ and Q$%
_{2}$, respectively, that are unambiguosly related to formation of the
relevant even-membered cycle. It deserves recalling here that the
odd-membered circuits R$_{1}$ and R$_{2}$ are represented by significant
stabilizing corrections $\mathcal{E}_{(3)}(R_{1})$ and\ $\mathcal{E}%
_{(5)}(R_{2}),$ respectively, that also are related to cyclization [17]. A
unified viewpoint to these two conclusions allows us to formulate the
following rule: A conjugated circuit (cycle) containing $k$ C=C and $k$ C$-$%
C bonds alternately is described by a decisive increment of the $k$th order (%
$\mathcal{E}_{(k)}^{(cycl)}$) within the power series for the relevant total
energy, the sign of which is determined by the parity factor $(-1)^{k+1}$\
(provided that a negative energy unit is used). The above-formulated
statement evidently is nothing more than a perturbative analogue of the H%
\"{u}ckel ($4n+2/4n$) rule [1, 2]. A similar result has been obtained
earlier when studying pericyclic reactions [49].

The above-obtained zero values of third order energies of the even-membered
circuits Q$_{n}$ (n=1,2..) support our assumption that these circuits are
not among substructures responsible for distinct corrections $\mathcal{E}%
_{(3)}$\ of some Kekul\'{e} valence structures of biphenylene-like
hydrocarbons containing the same numbers of the simplest odd-membered
circuits R$_{1}$ (Sect. 3). Thus, substructures playing this role are sought
in the next subsection.

\begin{center}
\textbf{4.2. Monocycles with exocyclic methylene groups}
\end{center}

Let us start with 3,4-dimethylene cyclobutene denoted below by S$_{1}$ (Fig.
4). Arguments for such a choice are as follows: First, Kekul\'{e} valence
structures of Section 3 that were shown to be characterized by exceptionally
lowered third order energies (viz. I/II, II/I, II/II, etc.), contain a
circuit S$_{1}$ among their substructures. Second, the circuit S$_{1}$ is
likely to offer a triplet of first-neighboring C=C bonds necessary for
emergence of a non-zero third order energy (Sect. 2). Thus, let us consider
this circuit in a more detail.

The relevant principal matrices are as follows 
\begin{equation}
\mathbf{G}_{(1)}\mathbf{(}S_{1}\mathbf{)=-}\frac{\gamma }{4}\left\vert 
\begin{array}{ccc}
0 & 1 & 1 \\ 
-1 & 0 & -1 \\ 
-1 & 1 & 0%
\end{array}%
\right\vert ,\quad \mathbf{G}_{(2)}\mathbf{(}S_{1}\mathbf{)=-}\frac{\gamma
^{2}}{8}\left\vert 
\begin{array}{ccc}
0 & -1 & 0 \\ 
1 & 0 & 0 \\ 
0 & 0 & 0%
\end{array}%
\right\vert
\end{equation}%
and deserve comparison to matrices $\mathbf{G}_{(1)}\mathbf{(}R_{1}\mathbf{)}
$ and $\mathbf{G}_{(2)}\mathbf{(}R_{1}\mathbf{)}$\ [17] representing the
standard three-membered circuit R$_{1}$, viz. 
\begin{equation}
\mathbf{G}_{(1)}\mathbf{(}R_{1}\mathbf{)=-}\frac{\gamma }{4}\left\vert 
\begin{array}{ccc}
0 & 1 & -1 \\ 
-1 & 0 & 1 \\ 
1 & -1 & 0%
\end{array}%
\right\vert ,\quad \mathbf{G}_{(2)}\mathbf{(}R_{1}\mathbf{)=-}\frac{\gamma
^{2}}{8}\left\vert 
\begin{array}{ccc}
0 & 1 & -1 \\ 
-1 & 0 & 1 \\ 
1 & -1 & 0%
\end{array}%
\right\vert .
\end{equation}%
It is seen that BBOs and ABOs of all pairs of C=C bonds interact directly in
the structures both S$_{1}$ and R$_{1}$ and thereby triplets of
first-neighboring C=C bonds are contained there. Moreover, matrices $\mathbf{%
G}_{(1)}\mathbf{(}S_{1}\mathbf{)}$ and $\mathbf{G}_{(1)}\mathbf{(}R_{1}%
\mathbf{)}$\ are similar except for signs of some elements. Finally, the
same refers also to their products $\mathbf{G}_{(1)}\mathbf{G}_{(1)}^{+}%
\mathbf{(}S_{1}\mathbf{)}$ and $\mathbf{G}_{(1)}\mathbf{G}_{(1)}^{+}\mathbf{(%
}R_{1}\mathbf{).}$\ Consequences of these similarities are as follows%
\begin{equation}
\mathcal{E}_{(2)}(S_{1})=\mathcal{E}_{(2)}(R_{1})=\frac{3}{2}\gamma
^{2},\quad \mathcal{E}_{(4)}^{(-)}(S_{1})=\mathcal{E}_{(4)}^{(-)}(R_{1})=-%
\frac{18\gamma ^{4}}{64}.
\end{equation}%
Meanwhile, the matrix $\mathbf{G}_{(2)}\mathbf{(}S_{1}\mathbf{)}$\ differs
from $\mathbf{G}_{(2)}\mathbf{(}R_{1}\mathbf{)}$\ significantly. Indeed, two
non-zero elements only are present in the former matrix in contrast to the
latter, namely the elements $\mathbf{G}_{(2)12}\mathbf{(}S_{1}\mathbf{)}$\
and $\mathbf{G}_{(2)21}\mathbf{(}S_{1}\mathbf{)}$\ representing the indirect
interactions of orbitals of exocyclic bonds C$_{1}$=C$_{4}$ and C$_{2}$=C$%
_{5}$. As a result, the stabilizing component of the fourth order energy $%
\mathcal{E}_{(4)}^{(+)}(S_{1})$\ also is accordingly lower as compared to $%
\mathcal{E}_{(4)}^{(+)}(R_{1})$\ as exhibited below in Eq.(37). Another
distinctive feature of the circuit S$_{1}$ consists in opposite signs of
elements of matrices $\mathbf{G}_{(2)}\mathbf{(}S_{1}\mathbf{)}$\ and $%
\mathbf{G}_{(1)}\mathbf{(}S_{1}\mathbf{)}$\ contributing to the correction $%
\mathcal{E}_{(3)}(S_{1})$ and thereby in negative signs of products\ $%
\mathbf{G}_{(2)12}\mathbf{(}S_{1}\mathbf{)\cdot G}_{(1)12}\mathbf{(}S_{1}%
\mathbf{)}$\ and $\mathbf{G}_{(2)21}\mathbf{(}S_{1}\mathbf{)\cdot G}_{(1)21}%
\mathbf{(}S_{1}\mathbf{)}$. Consequently, the third order energy of the
circuit S$_{1}$ also is a negative quantity (see Eq.(37)). Since the
decisive indirect interactions $\mathbf{G}_{(2)12}\mathbf{(}S_{1}\mathbf{)}$%
\ and $\mathbf{G}_{(2)21}\mathbf{(}S_{1}\mathbf{)}$\ are mediated here by
BOs of the endocyclic (C$_{3}$=C$_{6}$) bond, the negative sign of $\mathcal{%
E}_{(3)}(S_{1})$ may be also traced back to the destabilizing mediating
effect of BOs of this particular bond. By contrast, all products of matrix
elements contained in the expression for $\mathcal{E}_{(3)}(R_{1})$ are of
positive signs and this consequently refers to the correction itself. Thus,
the circuit R$_{1}$ offers us an example of the positive mediating effect of
BOs of the remaining bond (e.g. of C$_{3}$=C$_{6}$)\ in the indirect
interactions between those of any pair of first-neighboring C=C bonds (i.e.
of C$_{1}$=C$_{4}$ and C$_{2}$=C$_{5}$) [Distinction between a positive
mediating effect and a negative one proves to be especially useful if the
structure under study contains a combination of both circuits R$_{1}$ and S$%
_{1}$ (Section 5)]. Let us now collect\ the above-discussed energy increments%
\begin{eqnarray}
\mathcal{E}_{(3)}(S_{1}) &=&-\frac{1}{4}\gamma ^{3},\quad \mathcal{E}%
_{(4)}^{(+)}(S_{1})=\frac{8\gamma ^{4}}{64},\quad \mathcal{E}%
_{(4)}^{{}}(S_{1})=-\frac{10\gamma ^{4}}{64},  \notag \\
\mathcal{E}_{(3)}(R_{1}) &=&\frac{3}{4}\gamma ^{3},\quad \mathcal{E}%
_{(4)}^{(+)}(R_{1})=\frac{24\gamma ^{4}}{64},\quad \mathcal{E}%
_{(4)}^{{}}(R_{1})=\frac{6\gamma ^{4}}{64}
\end{eqnarray}%
and note that the corrections $\mathcal{E}_{(3)}(S_{1})$\ and $\mathcal{E}%
_{(3)}(R_{1}),$\ as well as $\mathcal{E}_{(4)}(S_{1})$\ and $\mathcal{E}%
_{(4)}(R_{1})$ are of comparable absolute values. This allows us to expect
that the circuit(s) S$_{1}$ (if any) participate(s) in the formation of both
third and fourth order energies of the relevant Kekul\'{e} valence
structures of biphenylene-like hydrocarbons.

The four-membered circuit S$_{2}$\ containing four exocyclic methylenes
(Fig. 4) also deserves our attention. As with the above-considered
four-membered circuit Q$_{2},$ the new circuit S$_{2}$ also is characterized
by a zero second order matrix $\mathbf{G}_{(2)}\mathbf{(}S_{2}\mathbf{).}$
Interpretation of this result is similar to that referring to the former
circuit Q$_{2}$. The relevant energy corrections are then as follows\ 
\begin{equation}
\mathcal{E}_{(2)}(S_{2})=\mathcal{E}_{(2)}(Q_{2})=2\gamma ^{2},\quad 
\mathcal{E}_{(3)}(S_{2})=0,\quad \mathcal{E}_{(4)}(S_{2})=\mathcal{E}%
_{(4)}(Q_{2})=-\frac{32\gamma ^{4}}{64}
\end{equation}%
and coincide with those of Q$_{2}$ to within fourth order terms inclusive
(see Eqs.(31) and (32)). This implies that the circuit S$_{2}$ (if any) is
able to make a negative contribution to the fourth order energy of the given
Kekul\'{e} valence structure.

The substructures S$_{1}$ and S$_{2}$ actually are present in certain
combinations with the standard CCs R$_{n}$ and Q$_{n}$ ($n=1,2...$) in the
individual Kekul\'{e} valence structures of Section 3. In this connection,
we will consider some typical examples of "cooperation" between circuits of
different series in making up separate energy corrections in the next
Section.

\begin{center}
\textbf{5. Increments of individual conjugated circuits to stabilities of
some selected Kekul\'{e} valence structures}
\end{center}

Simplest combinations of CCs of different series arise in the Kekul\'{e}
valence structures V/I, V/II and V/III\ of benzocyclobutene V (Fig. 5), all
of them containing four C=C bonds ($N=4$) as it was the case with Q$_{2}$.
Meanwhile, the relevant number of C$-$C bonds is now higher ($N\prime =5$).
The usual relation between the second order energy ($\mathcal{E}_{(2)}$) and
the parameter $N\prime $ (Sect. 2) yields $5\gamma ^{2}/2$. This value is
referred to below as the anticipated one.

\begin{figure}[!ht]
	\begin{center}
	\vspace*{5 pt}
	\includegraphics[width=12cm]{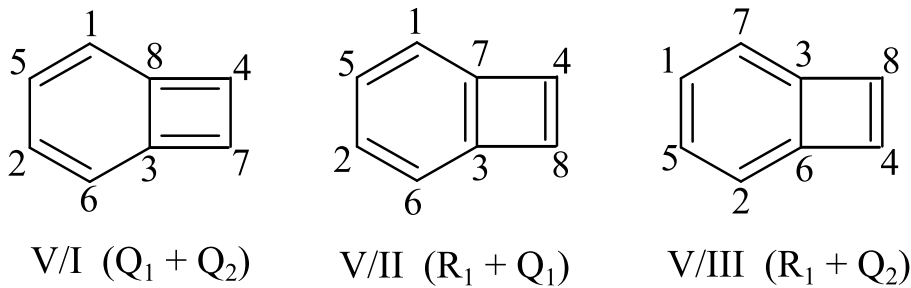}
  \caption{ Kekul\'{e} valence structures of benzocyclobutene (V) along with
their compositions in terms of standard conjugated circuits R$_{n}$ and Q$%
_{n},n=1,2,...$ . Numbers of carbon atoms and/or of their 2p$_{z}$ AOs also
are shown.}
	\label{fig5}
	\end{center}
\end{figure}	

Let us start with the structure V/I containing CCs of the Q$_{n}$ series
only, viz. Q$_{1}$ and Q$_{2}$. This structure may be also alternatively
regarded as a "perturbed" circuit Q$_{2},$ where the "perturbation" consists
in the emergence of an additional bond C$_{3}-$C$_{8}$ and thereby in the
formation of the additional circuit Q$_{1}$. The overall set of energy
increments is then as follows \ 
\begin{equation}
\mathcal{E}_{(2)}(V/I)=\frac{3}{2}\gamma ^{2},\quad \mathcal{E}%
_{(3)}(V/I)=0,\quad \mathcal{E}_{(4)}(V/I)=\mathcal{E}_{(4)}^{(-)}(V/I)=-%
\frac{14\gamma ^{4}}{64}.
\end{equation}%
To comment these results, let us invoke the relevant principal matrices. The
first order matrix $\mathbf{G}_{(1)}\mathbf{(}V/I\mathbf{)}$\ and its
derivative $\mathbf{G}_{(1)}\mathbf{G}_{(1)}^{+}\mathbf{(}V/I\mathbf{)}$\
take the forms 
\begin{equation}
\mathbf{G}_{(1)}\mathbf{(}V/I\mathbf{)=-}\frac{\gamma }{4}\left\vert 
\begin{array}{cccc}
0 & 1 & 0 & -1 \\ 
-1 & 0 & 1 & 0 \\ 
0 & -1 & 0 & 0 \\ 
1 & 0 & 0 & 0%
\end{array}%
\right\vert ,\mathbf{G}_{(1)}\mathbf{G}_{(1)}^{+}\mathbf{(}V/I\mathbf{)=}%
\frac{\gamma ^{2}}{16}\left\vert 
\begin{array}{cccc}
2 & 0 & -1 & 0 \\ 
0 & 2 & 0 & -1 \\ 
-1 & 0 & 1 & 0 \\ 
0 & -1 & 0 & 1%
\end{array}%
\right\vert .
\end{equation}%
It is seen that the matrix $\mathbf{G}_{(1)}\mathbf{(}V/I\mathbf{)}$\
contains zero elements in the positions 3,4 and 4,3 referring to the circuit
Q$_{1}$. This fact is in line with the vanishing matrix $\mathbf{G}_{(1)}%
\mathbf{(}Q_{1}\mathbf{)}$\ (see Eq.(22)). Comparison of the matrix $\mathbf{%
G}_{(1)}\mathbf{(}V/I\mathbf{)}$\ to $\mathbf{G}_{(1)}\mathbf{(}Q_{2}\mathbf{%
)}$\ of Eq.(30), in turn, shows that the total number of non-zero elements
becomes lower due to the above-described "perturbation". The relevant second
order energy $\mathcal{E}_{(2)}(V/I)$\ also is accordingly decreased by $%
\gamma ^{2}/2$\ as compared to $\mathcal{E}_{(2)}(Q_{2})$\ of Eq.(31).
Moreover, the difference between the actual value of $\mathcal{E}_{(2)}(V/I)$%
\ and the anticipated one ($5\gamma ^{2}/2$) equals to $\gamma ^{2}$ and
thereby coincides with the relevant (negative) cyclization energy $\mathcal{E%
}_{(2)}^{(c)}(Q_{1})$\ of Eq.(29). Hence, a second order destabilization may
be concluded to take place in the structure V/I that is entirely due to the
presence of the two-membered circuit Q$_{1}$.\ 

As opposed to matrices of Eq.(40), the remaining matrix $\mathbf{G}_{(2)}%
\mathbf{(}V/I\mathbf{)}$\ vanishes as it was the case with $\mathbf{G}_{(2)}%
\mathbf{(}Q_{2}\mathbf{)}$\ (Subsect. 4.1). As a result, the third order
energy $\mathcal{E}_{(3)}(V/I)$\ also takes a zero value. This fact is in
line with vanishing corrections $\mathcal{E}_{(3)}(Q_{1})$\ and $\mathcal{E}%
_{(3)}(Q_{2})$\ (Subsect. 4.1) and causes no surprise. A zero stabilizing
component ($\mathcal{E}_{(4)}^{(+)}(V/I)$)\ of the fourth order energy is
another implication of the equality\ $\mathbf{G}_{(2)}\mathbf{(}V/I\mathbf{%
)=0}$.\ Thus, the systems V/I and Q$_{2}$ are similar in respect of
vanishing fourth order stabilization too. By contrast, the relevant
destabilizing components differ one from another significantly as comparison
of Eqs.(31) and (39) shows, viz. the absolute value of $\mathcal{E}%
_{(4)}^{(-)}(V/I)$\ coincides with almost a half of that of $\mathcal{E}%
_{(4)}^{(-)}(Q_{2}).$\ This is because the zero elements $\mathbf{G}_{(1)34}$%
\ and $\mathbf{G}_{(1)43}$\ determine lower absolute values of those of the
matrix $\mathbf{G}_{(1)}\mathbf{G}_{(1)}^{+}\mathbf{(}V/I\mathbf{)}$\ as
compared to $\mathbf{G}_{(1)}\mathbf{G}_{(1)}^{+}\mathbf{(}Q_{2}\mathbf{)}$\
(see Eqs.(30) and (40)). Consequently, emergence of the additional circuit Q$%
_{1}$ contributes to suppression of the fourth order destabilization caused
by the circuit Q$_{2}$ in the structure V/I.\ 

Let us now turn to the structure V/II containing CCs of both $4n+2$ and $4n$
series, namely R$_{1}$ and Q$_{1}$. The relevant matrices $\mathbf{G}_{(1)}%
\mathbf{(}V/II\mathbf{)}$\ and $\mathbf{G}_{(2)}\mathbf{(}V/II\mathbf{)}$\
are as follows%
\begin{equation}
\mathbf{G}_{(1)}\mathbf{(}V/II\mathbf{)=-}\frac{\gamma }{4}\left\vert 
\begin{array}{cccc}
0 & 1 & -1 & 0 \\ 
-1 & 0 & 1 & 0 \\ 
1 & -1 & 0 & 0 \\ 
0 & 0 & 0 & 0%
\end{array}%
\right\vert ,\mathbf{G}_{(2)}\mathbf{(}V/II\mathbf{)=-}\frac{\gamma ^{2}}{8}%
\left\vert 
\begin{array}{cccc}
0 & 1 & -1 & 1 \\ 
-1 & 0 & 1 & -1 \\ 
1 & -1 & 0 & 0 \\ 
-1 & 1 & 0 & 0%
\end{array}%
\right\vert .
\end{equation}%
As with the former first order matrix $\mathbf{G}_{(1)}\mathbf{(}V/I\mathbf{%
),}$\ the new one ($\mathbf{G}_{(1)}\mathbf{(}V/II\mathbf{)}$)\ also
contains zero elements in the positions 3,4 and 4,3 referring to the
two-membered circuit Q$_{1}$. Moreover, total numbers of non-zero elements
are uniform in both matrices $\mathbf{G}_{(1)}\mathbf{(}V/I\mathbf{)}$\ and $%
\mathbf{G}_{(1)}\mathbf{(}V/II\mathbf{)}$\ along with the consequent second
order energies (see Eqs. (39) and (42)). Finally, the energy $\mathcal{E}%
_{(2)}(V/II)$\ is lowered by $\gamma ^{2}$ as compared to the anticipated
value ($5\gamma ^{2}/2$) as previously. Thus, a second order destabilization
may be concluded to take place in the structure V/II too that may be traced
back to the presence of the circuit Q$_{1}$. Besides, an analogous state of
things refers also to Q$_{1}$-containing structures of Section 3.

Let us return again to Eq.(41) and note that the non-zero part (block) of
the matrix $\mathbf{G}_{(1)}\mathbf{(}V/II\mathbf{)}$\ resembles that of the
circuit R$_{1}$ (see Eq. (35)) in accordance with the expectation. The same
then accordingly refers to the non-zero block of the product $\mathbf{G}%
_{(1)}\mathbf{G}_{(1)}^{+}\mathbf{(}V/II)$. Implications of these
similarities are as follows%
\begin{equation}
\mathcal{E}_{(2)}(V/II)=\mathcal{E}_{(2)}(R_{1})=\frac{3}{2}\gamma
^{2},\quad \mathcal{E}_{(4)}^{(-)}(V/II)=\mathcal{E}_{(4)}^{(-)}(R_{1})=-%
\frac{18\gamma ^{4}}{64}.
\end{equation}%
By contrast, the second order matrix $\mathbf{G}_{(2)}\mathbf{(}V/II\mathbf{)%
}$\ of Eq.(41) contains non-zero "intercircuit" elements in the positions
1,4[4,1] and 2,4[4,2]. These elements represent new stabilizing indirect
interactions that emerge in the structure V/II. Although these extra
elements exert no influence upon the third order energy [because of zero
products like $\mathbf{G}_{(2)14}\cdot \mathbf{G}_{(1)14}]$\ and the
correction $\mathcal{E}_{(3)}(V/II)$ keeps to coincide with $\mathcal{E}%
_{(3)}(R_{1})$\ (see Eq.(43)), the stabilizing fourth order energy component 
$\mathcal{E}_{(4)}^{(+)}(V/II)$\ grows significantly vs. $\mathcal{E}%
_{(4)}^{(+)}(R_{1})=24\gamma ^{4}/64$\ [17]. The same then accordingly
refers to total values of fourth order energies. We therefore obtain 
\begin{equation}
\mathcal{E}_{(3)}(V/II)=\mathcal{E}_{(3)}(R_{1})=\frac{3}{4}\gamma
^{3},\quad \mathcal{E}_{(4)}^{(+)}(V/II)=\frac{40\gamma ^{4}}{64},\quad 
\mathcal{E}_{(4)}(V/II)=\frac{22\gamma ^{4}}{64}.
\end{equation}%
Hence, the high (positive) value of the fourth order energy of the structure
V/II is entirely due to an additional stabilizing increment of the
intercircuit origin. The relatively high fourth order energies of the R$_{1}$%
+Q$_{1}$-containing Kekul\'{e} valence structures of biphenylene-like
hydrocarbons (Section 3) also may be rationalized similarly.

Let us dwell now on the last structure of benzocyclobutene V/III, containing
the standard circuits R$_{1}$ and Q$_{2}.$\ The relevant principal matrices
are as follows\ \ 
\begin{equation*}
\mathbf{G}_{(1)}\mathbf{(}V/III\mathbf{){}=-{}}\frac{\gamma }{4}\left\vert 
\begin{array}{cccc}
0 & 1 & -1 & 0 \\ 
-1 & 0 & 1 & 1 \\ 
1 & -1 & 0 & -1 \\ 
0 & -1 & 1 & 0%
\end{array}%
\right\vert ,\mathbf{G}_{(2)}\mathbf{(}V/III\mathbf{)=-}\frac{\gamma ^{2}}{8}%
\left\vert 
\begin{array}{cccc}
0 & 1 & -1 & 1 \\ 
-1 & 0 & 0 & 0 \\ 
1 & 0 & 0 & 0 \\ 
-1 & 0 & 0 & 0%
\end{array}%
\right\vert
\end{equation*}%
\begin{eqnarray}
&&  \notag \\
\mathbf{G}_{(1)}\mathbf{G}_{(1)}^{+}\mathbf{(}V/III\mathbf{)} &\mathbf{=}&%
\frac{\gamma ^{2}}{16}\left\vert 
\begin{array}{cccc}
2 & -1 & -1 & -2 \\ 
-1 & 3 & -2 & 1 \\ 
-1 & -2 & 3 & 1 \\ 
-2 & 1 & 1 & 2%
\end{array}%
\right\vert ,
\end{eqnarray}%
whereas the energy corrections concerned take the form \ 
\begin{eqnarray}
\mathcal{E}_{(2)}(V/III) &=&\frac{5}{2}\gamma ^{2},\quad \mathcal{E}%
_{(3)}(V/III)=\frac{2}{4}\gamma ^{3},\quad \mathcal{E}_{(4)}^{(+)}(V/III)=%
\frac{16\gamma ^{4}}{64},  \notag \\
\mathcal{E}_{(4)}^{(-)}(V/III) &=&-\frac{50\gamma ^{4}}{64},\quad \mathcal{E}%
_{(4)}(V/III)=-\frac{34\gamma ^{4}}{64}.
\end{eqnarray}%
It is seen that the number of non-zero elements of the matrix $\mathbf{G}%
_{(1)}\mathbf{(}V/III\mathbf{)}$ coincides with the relevant two-fold actual
number of C$-$C bonds ($2N\prime $) in contrast to the former matrices $%
\mathbf{G}_{(1)}\mathbf{(}V/II\mathbf{)}$\ and $\mathbf{G}_{(1)}\mathbf{(}V/I%
\mathbf{).}$\ The resulting second order energy $\mathcal{E}_{(2)}(V/III)$\
also accordingly equals to the anticipated value ($5\gamma ^{2}/2$). This
implies that the structure V/III is the most stable one in the case of
benzocyclobutene (V), the overall order of stability being V/III>V/II>V/I. 
Besides, the structure V/III also may be regarded as
a "perturbed" circuit Q$_{2}$. In contrast to the former case of V/I,
however, formation of the new bond C$_{3}$-C$_{6}$ is now accompanied by
emergence of an additional pair of C=C bonds, the BOs of which interact
directly (viz. C$_{2}$=C$_{6}$ and C$_{3}$=C$_{7}$), and, consequently, the
relevant second order energy is higher as compared to $\mathcal{E}%
_{(2)}(Q_{2}).$

The second order matrix $\mathbf{G}_{(2)}\mathbf{(}V/III\mathbf{)}$\ of
Eq.(44) exhibits even more surprising properties. Indeed, this matrix
contains zero elements in the positions 2,3 and 3,2 so that no submatrix
(block) like $\mathbf{G}_{(2)}\mathbf{(}R_{1}\mathbf{)}$\ arises there in
spite of the presence of the circuit R$_{1}$ in the structure concerned. A
more detailed analysis of expressions for elements $\mathbf{G}%
_{(2)23}(V/III) $ shows that mediating increments of BOs of bonds C$_{1}$=C$%
_{5}$ and C$_{4}$=C$_{8}$ are of the same absolute values and of opposite
signs and thereby cancel out one another. This result causes little surprise
if we recall (i) additivity of any second order element $\mathbf{G}_{(2)il}$%
\ with respect to mediators seen from Eq.(8) and (ii) opposite signs of
mediating effects of BOs of endocyclic C=C bonds in the isolated circuits R$%
_{1}$ and S$_{1}$ (Subsect. 4.2). The above-discussed zero elements $\mathbf{%
G}_{(2)23}(V/III), $\ in turn, imply vanishing products like $\mathbf{G}%
_{(2)23}\mathbf{G}_{(1)23}$ in the final formula for the third order energy $%
\mathcal{E}_{(3)}(V/III).$ The actual value of the latter then coincides
with $2\gamma ^{3}/4$ instead of $3\gamma ^{3}/4$ representing the circuit R$%
_{1}$ (see Eq. (37)). Moreover, the energy concerned proves to be additive
with respect to transferable increments of the circuits R$_{1}$ and S$_{1},$
viz. \ \ 
\begin{equation}
\mathcal{E}_{(3)}(V/III)=\mathcal{E}_{(3)}(R_{1})+\mathcal{E}_{(3)}(S_{1}),
\end{equation}%
where components of the right-hand side are of opposite signs. Hence, the
circuit S$_{1}$ undoubtedly is among substructures determining the third
order energy of the most stable Kekul\'{e} valence structure of
benzocyclobutene (V/III). Finally, the above-outlined scheme of formation is
easily transferable to third order energies of S$_{1}$-containing structures
of both biphenylene (I/II) and related hydrocarbons (II/II, III/I, etc.).

The fourth order energy $\mathcal{E}_{(4)}(V/III)$\ of Eq.(45) also deserves
some attention. As with the former structures V/I and V/II, the overall
relation between this energy and the CCs contained is much less
straightforward as compared to the above-discussed simple relations for $%
\mathcal{E}_{(2)}(V/III)$\ and $\mathcal{E}_{(3)}(V/III)$. Nevertheless,
there are arguments for a conclusion that the non-standard circuit S$_{1}$
participates in the formation of the correction $\mathcal{E}_{(4)}(V/III)$\
too: First, the actual value of this correction ($-34\gamma ^{4}/64$) is
considerably closer to the sum of increments of three circuits (R$_{1}$, Q$%
_{2}$ and S$_{1}$) rather than of two ones (R$_{1}$ and Q$_{2}$) [these sums
correspondingly coincide with $-36\gamma ^{4}/64$\ and $-26\gamma ^{4}/64$].
The second argument follows from comparison of fourth order energies of all
Kekul\'{e} valence structures of benzocyclobutene. Indeed, the relative
values of these corrections [i.e. $\mathcal{E}_{(4)}(V/II)>\mathcal{E}%
_{(4)}(V/I)>\mathcal{E}_{(4)}(V/III)$] are consistent with the increasing
total number of "destabilizing" contributors, namely with zero, one (i.e. Q$%
_{2}$) and two (i.e. Q$_{2}$ and S$_{1}$), respectively.

As already mentioned, the four-membered circuit S$_{2}$ (Fig. 4) also is
able to contribute to fourth order energies of the relevant Kekul\'{e}
valence structures. The most stable structure of biphenylene (I/I) (Section
3) offers us an example, wherein a circuit S$_{2}$ is present in combination
with the standard ones (2R$_{1}$+Q$_{3}$). Comparison of the actual (i.e. of
zero) value of the correction $\mathcal{E}_{(4)}(I/I)$\ (see Eq.(11)) to
anticipated values following from different additive schemes support our
expectation about importance of the circuit S$_{2}$: Summing up the
increments of the standard circuits only [i.e. of $2\mathcal{E}_{(4)}(R_{1})$
and $\mathcal{E}_{(4)}(Q_{3})$] yields $24\gamma ^{4}/64$\ (see Eqs.(33) and
(37)), whilst a subsequent addition of $\mathcal{E}_{(4)}(S_{2})$\ of
Eq.(38) results into $-8\gamma ^{4}/64,$ the latter outcome being much
closer to the actual zero value. To strengthen these arguments, let us
consider the principal matrices of the structure I/I in a more detail.

The first order matrix $\mathbf{G}_{(1)}\mathbf{(}I/I\mathbf{)}$\ contains
two 3$\times $3 -dimensional blocks in its diagonal positions that represent
individual circuits R$_{1}$ and coincide with the matrix $\mathbf{G}_{(1)}%
\mathbf{(}R_{1}\mathbf{)}$\ of Eq.(35) in accordance with the expectation.
So far as the off-diagonal blocks of the same matrix are concerned, non-zero
elements (either 1 or $-$1) stand here in the positions 1,6[6,1] and
3,4[4,3]. These "intercircuit" elements determine an increase of the second
order energy of the structure I/I as compared to $2\mathcal{E}_{(2)}(R_{1})$%
\ (see Eqs.(9) and (36)). Further, the second order matrix of the structure
I/I is as follows\ \ 
\begin{equation}
\mathbf{G}_{(2)}\mathbf{(}I/I\mathbf{)=-}\frac{\gamma ^{2}}{8}\left\vert 
\begin{array}{cccccc}
0 & -1 & 1 & 0 & -1 & 0 \\ 
1 & 0 & -1 & 1 & 0 & -1 \\ 
-1 & 1 & 0 & 0 & 1 & 0 \\ 
0 & -1 & 0 & 0 & 1 & -1 \\ 
1 & 0 & -1 & -1 & 0 & 1 \\ 
0 & 1 & 0 & 1 & -1 & 0%
\end{array}%
\right\vert
\end{equation}%
and accordingly contains two submatrices like $\mathbf{G}_{(2)}\mathbf{(}%
R_{1})$\ of Eq.(35) in its principal diagonale. Since\ non-zero elements
take distinct positions in the off-diagonal blocks of matrices $\mathbf{G}%
_{(2)}\mathbf{(}I/I\mathbf{)}$ and $\mathbf{G}_{(1)}\mathbf{(}I/I\mathbf{)}$%
, these blocks make no contributions to the third order energy $\mathcal{E}%
_{(3)}(I/I)$\ and the latter coincides with the two-fold increment of an
individual circuit R$_{1}$ (see Eqs.(10) and (37)). More importantly, the
matrix $\mathbf{G}_{(2)}\mathbf{(}I/I\mathbf{)}$\ contains zero elements in
the positions 1,4[4,1], 1,6[6,1], 3,4[4,3] and 3,6[6,3] referring to the
substructure S$_{2}$ (Fig. 1) [Elements $\mathbf{G}_{(2)13}\mathbf{(}I/I%
\mathbf{)}$\ and $\mathbf{G}_{(2)46}\mathbf{(}I/I\mathbf{)}$\ (along with
their counterparts in the positions 3,1 and 6,4, respectively) make an
exception owing to additional mediating effects of the second and fifth C=C
bonds]. Consequently, the total number of significant elements is relatively
low in the matrix $\mathbf{G}_{(2)}\mathbf{(}I/I\mathbf{).}$\ The same
evidently refers to the stabilizing component of the fourth order energy $%
\mathcal{E}_{(4)}^{(+)}(I/I)$\ and thereby to the total correction $\mathcal{%
E}_{(4)}(I/I)$ itself. Hence, participation of the circuit (substructure) S$%
_{2}$ in the formation of the fourth order energy $\mathcal{E}_{(4)}(I/I)$\
is beyond any doubt.

\begin{center}
\textbf{6. Conclusions}
\end{center}

The principal achievement of the present study consists in revealing how
various aspects of constitution of individual Kekul\'{e} valence structures
of biphenylene-like hydrocarbons are reflected in their total pi-electron
energies and thereby in relative stabilities. Sseparate members of the power
series for these energies are shown to be governed by the following rules:

1) Zero order energies are uniform for all Kekul\'{e} valence structures of
the same hydrocarbon, whereas the relevant first order corrections vanish.

2) Second order energies of the Q$_{1}$-containing Kekul\'{e} valence
structures are lowered by $M\gamma ^{2}$\ as compared to those of the
remaining (Q$_{1}$ -free) structures, where $M$ stands for the number of the
circuits Q$_{1}$. This decrease is unambiguosly traced back to the
destabilizing influence of just these simplest even-membered circuits.

3) Third order energies of Kekul\'{e} valence structures of biphenylene-like
hydrocarbons ($\mathcal{E}_{(3)}$) are additive quantities with respect to
transferable increments of the standard three-membered CCs R$_{1}$ and of
the newly-introduced circuits S$_{1}$, correspondingly coinciding with $%
3\gamma ^{3}/4$\ and $-\gamma ^{3}/4.$\ As a result, the circuit(s) S$_{1}$
(if any) exert(s) destabilizing influence(s) upon the third order energies
of the relevant Kekul\'{e} valence structures. These energies are then
exceptionally reduced vs. the anticipated values following from the relation 
$\mathcal{\varepsilon }_{(3)}=3\gamma ^{3}K/4$\ established previously for
benzenoid hydrocarbons ($K$ stands here for the number of the standard
circuits R$_{1}$).

4) Fourth order energies of the Q$_{2}$-containing Kekul\'{e} valence
structures take relatively low (negative) values mainly because of the
destabilizing influence just of the four-membered conjugated circuit(s) Q$%
_{2}$.

As opposed to the simple and exact form of the first three rules, the last
one is of an approximate nature only. This is because presence (or absence)
of the circuit Q$_{2}$ is not the only factor determining the actual value
of the fourth order correction and other peculiarities of the Kekul\'{e}
valence structure concerned also are able to play an important role here. In
this respect, two additional rules may be added here:

5) Kekul\'{e} valence structures of biphenylene-like hydrocarbons containing
the simplest standard circuits R$_{1}$ and Q$_{1}$ possessing a single
common C=C bond are characterized by an excessive fourth order stabilization
of the intercircuit origin.

6) The newly-introduced circuits with two and four exocyclic C=C bonds (S$%
_{1}$ and S$_{2}$) also participate in the formation of the fourth order
energies of the relevant Kekul\'{e} valence structures. Moreover, both
circuits S$_{1}$ and S$_{2}$ contribute to the fourth order destabilization
of the given structure along with the standard circuit(s) Q$_{2}.$

In summary, the above-formulated rules corroborate our principal hypothesis
(Sect 3) about the decisive role of even-membered conjugated circuits Q$_{1}$
and Q$_{2}$\ in the second and fourth order destabilization, respectively,
of the relevant Kekul\'{e} valence structures of biphenylene-like
hydrocarbons, as well as indicate the substructures (circuits) S$_{1}$ and S$%
_{2}$ (if any) to participate in the formation of the third and fourth order
energies along with the standard circuits of the $4n+2$ and $4n$ series (R$%
_{n}$ and Q$_{n},n=1,2...$).

In respect of comparison of the present perturbative approach to the CC
theory, the principal conclusions are as follows:

i) Total energies of individual Kekul\'{e} valence structures of
biphenylene-like hydrocarbons generally differ one from another in terms of
the second order of the power series (as the second rule indicates).
Meanwhile, the analogous distinctions were shown to be of the third order
magnitude in the case of benzenoid hydrocarbons [17] containing the circuits
of the $4n+2$ series only. This implies that an assumption about uniform
"weights" of all Kekul\'{e} valence structures is less justified for
biphenylene-like hydrocarbons as compared to benzenoids. Such a conclusion,
in turn, serves as a deductive accounting for less satisfactory results of
the simplest version of the CC theory [based on such an assumtion] in the
case of non-benzenoid hydrocarbons (Section 1).

ii) The even(odd)-membered circuits Q$_{n}$(R$_{n}$), $n=1,2...$ contribute
to energy corrections of even (odd) orders as the above-obtained results
indicate so that participation of the former (latter) actually starts with
second (third) order terms of the power series. This implies that
destabilizing increments of the even-membered circuits Q$_{n}(n=1,2,..)$
generally are more important as compared to stabilizing contributions of the
odd-membered ones (R$_{n}$). Accordingly, the absolute values of parameters
of CC models representing the circuits Q$_{n}$ should exceed those referring
to R$_{n}$ for the same $n$ value. Parameters of Ref.[14] comply with this
recommendation.

iii) The approach applied allows the Kekul\'{e} valence structures of a
certain bipheny lene- like hydrocarbon to be ordered according to their
total pi-electron energies and thereby relative stabilities. Thus, the
present study offers an independent deductive criterion for evaluation of
relative importances of individual structures. In particular, the S$_{2}$%
-containing structures (if any) are expected to be the most important ones
[e.g. the structures I/I and IV/I of biphenylene (I) and bent [3] phenylene
(IV), respectively], whereas the S$_{1}$-containing ones take the second
place. [This result may be traced back to absence of "destabilizing"
even-membered circuits Q$_{1}$ and Q$_{2}$ in the structures concerned].
Meanwhile, lowest relative stabilities and/or importances are predicted for
Kekul\'{e} valence structures in which either the neighboring hexagonal
rings are connected by two C=C bonds (such as I/IV) or two exocyclic C=C
bonds are attached to a certain hexagonal ring (e. g. II/IV, III/V etc.).
Thus, excluding of these particular structures suggested in Ref.[33] is
supported by the results of the perturbative approach.

iv) The above-established participation of the newly-introduced circuits S$%
_{1}$ and S$_{2}$ (Fig. 4) in the formation of total energies and thereby
relative stabilities of the relevant Kekul\'{e} valence structures of
biphenylene-like hydrocarbons indicates that these substructures should be
considered as supplementary conjugated circuits for these compounds and
thereby should be incorporated into the relevant CC model(s).

.

\begin{center}
\textbf{REFERENCES}
\end{center}

[1] C. A. Coulson, B. O'Leary and R. B. Mallion, \textit{H\"{u}ckel Theory
for Organic Chemistry}, Acad. Press, London, 1978.

[2] K. Yates, \textit{H\"{u}ckel Molecular orbital Theory, }Acad. Press, New
York, 1978.

[3] G. W. Wheland, \textit{Resonance Theory in Organic Chemistry}, Wiley \&
Sons, New York, 1955.

[4] W. C. Herndon, Resonance Theory and the Enumeration of Kekul\'{e}
Structures, \textit{J. Chem. Educ}. \textbf{51} (1974) 10-15.

[5] M. Randic, On the Characterization of Local Aromatic Properties in
Benzenoid Hydrocarbons, \textit{Tetrahedron}, \textbf{30} (1974) 2067-2074.

[6] I. Gutman, On the Hall Rule in the Theory of Benzenoid Hydrocarbons, 
\textit{Int. J. Quant. Chem}. \textbf{74} (1999) 627-632.

[7] E. Clar, \textit{The Aromatic Sextet}, Wiley \& Sons, London, 1972.

[8] H. Hosoya, Clar's Aromatic Sextet and Sextet Polynomial, \textit{Topics
Curr. Chem}. \textbf{153} (1990) 255-272.

[9] W. C. Herndon and H. Hosoya, Parametric Valence Bond Calculations for
Benzenoid Hydrocarbons Using Clar Structures, \textit{Tetrahedron} \textbf{40%
} (1984) 3987-3995.

[10] I. Gutman, Cyclic Conjugation Effects in Polycyclic $\pi -$Electron
Systems, \textit{Monatsh. Chem}. \textbf{136} (2005) 1055-1069.

[11] M. Randic, Conjugated Circuits and Resonance Energies of Benzenoid
Hydrocarbons, \textit{Chem. Phys}. \textit{Lett.} \textbf{38} (1976) 68-70.

[12] M. Randic, A Graph Theoretical Approach to Conjugation and Resonance
Energies of Hydrocarbons, \textit{Tetrahedron} \textbf{33} (1977) 1905-1920.

[13] M. Randic, Aromaticity and Conjugation, \textit{J. Amer. Chem. Soc. }%
\textbf{99} (1977) 444-450.

[14] M. Randic, On the Role of Kekul\'{e} Valence Structures, \textit{Pure
Appl. Chem}. \textbf{55} (1983) 347-354.

[15] M. Randic, Aromaticity of Polycyclic Conjugated Hydrocarbons, \textit{%
Chem. Revs.} \textbf{103} (2003) 3449-3605.

[16] N. Trinajstic, \textit{Chemical Graph Theory}, CRC Press, Boca Raton
FL, 1983.

[17] V. Gineityte, Perturbative Analogue for the Concept of Conjugated
Circuits in Benzenoid Hydrocarbons, \textit{MATCH Commun. Math. Comput. Chem}%
. \textbf{72} (2014) 39-73.

[18] M. J. S. Dewar, R. C. Dougherty, \textit{The PMO Theory of Organic
Chemistry,} Plenum Press, New York, 1975.

[19] V. Gineityte, On the Relation between the Stabilization Energy of a
Molecular System and the Respective Charge Redistribution, \textit{J. Mol.
Struct. (Theochem) }\textbf{585} (2002) 15-25.

[20] V. Gineityte, An Alternative to the Standard PMO Theory of Conjugated
Hydrocarbons in Terms of Overlap Topologies, \textit{Int. J. Chem.} \textit{%
Model.} \textbf{5} (2013) 99-118.

[21] V. Gineityte, On Relative Stabilities of Distinct Polyenes. An
Extension of the Concept of Conjugated Paths, ArXiv: 1501.04734 (2015)

http://arxiv.org/abs/1501.04734

[22] N. Trinajstic, D. Plavsic, The Conjugated-Circuit Model Revisited, 
\textit{Croat. Chem. Acta} \textbf{62} (1989) 711-718.

[23] I. Gutman, S. J. Cyvin, Conjugated Circuits in Benzenoid Hydrocarbons, 
\textit{J. Mol. Struct. (Theochem)} \textbf{184} (1989) 159-163.

[24] D. J. Klein, N. Trinajstic, Foundations of Conjugated- Circuits Models, 
\textit{Pure Appl. Chem. }\textbf{61} (1989) 2107-2115.

[25] S. Nikolic, N. Trinajstic, D. J. Klein, The Conjugated-Circuits Model, 
\textit{J. Comput. Chem}. \textbf{14} (1990) 313-322

[26] D. Plavsic, S. Nikolic, N. Trinajstic, The Conjugated-Circuit Model:
The Optimum Parameters for Benzenoid Hydrocarbons, \textit{J. Math. Chem. }%
\textbf{8} (1991) 113-120.

[27] N. Trinajstic, N. Nikolic, D. J. Klein, Quantum-Mechanical and
Computational Aspects of the Conjugated Circuits Model, \textit{J. Mol.
Struct. (Theochem)} \textbf{229} (1991) 63-89.

[28] X. Guo, M. Randic, Recursive Method for Enumeration of Linearly
Independent and Minimal Conjugated Circuits in Benzenoid Hydrocarbons, 
\textit{J. Chem. Inf. Comput. Sci.} \textbf{34} (1994) 339-348.

[29] X. Guo, M. Randic, D. J. Klein, Analytical Expressions for the Count of
LM-Conjugated Circuits of Benzenoid Hydrocarbons, \textit{Int. J. Quant.
Chem.} \textbf{60} (1998) 943-958.

[30] C. D. Lin, Efficient Method for Calculating the Resonance Energy
Expression of Benzenoid Hydrocarbons Based on the Enumeration of Conjugated
Circuits, \textit{J. Chem. Inf. Comput. Sci}. \textbf{40} (2000) 778-783.

[31] X. Guo, M. Randic, Recursive Formulae for Enumeration of LM-Conjugated
Circuits in Structurally Related Benzenoid Hhydrocarbons, \textit{J. Math.
Chem.} \textbf{30} (2001) 325-342.

[32] T. Doslic, Counting Conjugated Circuits in Benzenoid Chains, \textit{%
MATCH Commun. Math. Comput. Chem.} \textbf{65} (2011) 775-784.

[33] M. Randic, A. T. Balaban, D. Plavsic, Applying the Conjugated Circuits
Method to Clar Structures of [n]Phenylenes for Determining Resonance
Energies, \textit{Phys. Chem. Chem. Phys.} \textbf{13} (2011) 20644-20648.

[34] M. Randic, D. Plavsic and N. Trinajstic, Maximum Valence Structures in
Nonbenzenoid Polycyclic Hydrocarbons, \textit{Struct. Chem}. \textbf{2}
(1991) 543-554.

[35] K. P. C. Vollhardt and D. L. Mohler, The Phenylenes: Synthesis,
Properties, and Reactivity. In: B. Halton (Ed.) \textit{Advances in Strain
in Organic Chemistry}, \textbf{5}, JAI Press, London.

[36] I. Fishtik, Biphenylene: Stabilization or Destabilization? \textit{J.
Phys. Org. Chem}. \textbf{24} (2011) 263-266.

[37] R. H. Mitchell and V. S. Iyer, How Aromatic Are the Benzene Rings in
Biphenylene? The Synthesis and NMR Properties of a Biphenyleno Fused
Dihydropyrene, \textit{J. Amer. Chem. Soc}. \textbf{118} (1996) 2903-2906.

[38] M. Randic, $\pi -$Electron Currents in Polycyclic Conjugated
Hydrocarbons of Decreasing Aromatic Character and a Novel Structural
Definition of Aromaticity, \textit{Open Org. Chem. J}. \textbf{5} (2011)
11-26.

[39] I. Gutman, A. Stajkovic, S. Markovic and P. Petrovic, Dependence of
Total $\pi -$ Electron Energy of Phenylenes on Kekul\'{e} Structure Count, 
\textit{J. Serb. Chem. Soc}. \textbf{59} (1994) 367-373.

[40] I. Gutman, S. Markovic, A. Stajkovic and S. Kamidzorac, Correlations
between $\pi -$ Electron Properties of Phenylenes and Their Hexagonal
Squeezes, \textit{J. Serb. Chem. Soc}. \textbf{61} (1996) 873-879.

[41] N. Trinajstic, The H\"{u}ckel Theory and Topology, In: G. A. Segal
(Ed.),\ \textit{Semiempirical Methods of Electronic Structure Calculations,
Part A, Techniques,} Plenum Press, New York and London, 1977.

[42] G. G. Hall, The Bond Orders of Alternant Hydrocarbon Molecules, \textit{%
Proc. Roy. Soc. (London)} \textbf{A229 }(1955) 251-259.

[43] V. Gineityte, Application of the Non-Canonical Method of Molecular
Orbitals for Investigation of Electronic Structures of Conjugated
Hydrocarbons, \textit{J. Mol. Struct. (Theochem),} \textbf{487} (1999)
231-240.

[44] M. J. S. Dewar and H. C. Longuet-Higgins, The Correspondence between
the Resonance and Molecular Orbital Theories, \textit{Proc. Roy. Soc}. 
\textbf{A214} (1952) 428-497.

[45] D. J. Klein and M. Randic, Innate Degree of Freedom of a Graph, \textit{%
J. Comput. Chem}. \textbf{8} (1987) 516-521.

[46] A. Graovac, I. Gutman, M. Randic and N. Trinajstic, Kekul\'{e} Index
for Valence Bond Structures of Conjugated Polycyclic Systems, \textit{J.
Amer. Chem. Soc.} \textbf{95} (1973) 6267-6273.

[47] A. Graovac, I. Gutman, M. Randic and N. Trinajstic, Kekul\'{e} Index
for Valence Bond Structures of Conjugated Systems Containing Cyclobutadiene, 
\textit{Collect. Czech. Chem. Commun}. \textbf{43} (1978) 1375-1392.

[48] Sh. El-Basil, Novel Graph-Theoretical Approach to Estimating the
Relative Importance of Individual Kekul\'{e} Valence Structures. III.
Nonalternate and Nonbenzenoid Hydrocarbons, \textit{Int. J. Quant. Chem.} 
\textbf{21} (1982) 793-797.

[49] V. Gineityte, A Simple Topological Factor Determining the Allowance of
Pericyclic Reactions, \textit{Int. J. Quant. Chem}. \textbf{108} (2008)
1141-1154.

[50] V. Gineityte, Overlap Topologies Determining the Predominant Routes of
Electrophilic and Nucleophilic Reactions, \textit{Int. J. Chem. Model.} 
\textbf{4} (2012) 189-211.

\end{document}